\documentstyle[12pt]{article}
\newlength{\extraspace}
\newlength{\extraspaces}
\newcommand{\ba}{\begin{eqnarray}
\addtolength{\abovedisplayskip}{\extraspaces}
\addtolength{\belowdisplayskip}{\extraspaces}
\addtolength{\abovedisplayshortskip}{\extraspace}
\addtolength{\belowdisplayshortskip}{\extraspace}}
\newcommand{\ea}{\end{eqnarray}}

\newcommand{\nonu}{\nonumber \\[.5mm]}
\newcommand{\A}{&\!\!\!}

\input{prepictex}
\input{pictex}
\input{postpictex}

\begin{document}

\thispagestyle{empty}

\hfill \parbox{3.5cm}{hep-th/ \\ SIT-LP-03/09}
\vspace*{1cm}

\begin{center}
{\large \bf New Einstein-Hilbert-type Action \\ 
and   \\
Superon-Graviton Model(SGM) of Nature}
\footnote{Based on a talk by K. Shima at International Conference on Mathematics and Nucler Physics 
for the 21st Century, 
March 8-13, 2003, Atomic Energy Authority, Cairo, Egypt}    \\[15mm]
{Kazunari SHIMA and Motomu Tsuda} \\[1mm]
{\em Laboratory of Physics, Saitama Institute of Technology} 
\footnote{e-mail:shima@sit.ac.jp, tsuda@sit.ac.jp}\\
{\em Okabe-machi, Saitama 369-0293, Japan}\\[2mm]
{Manabu Sawaguchi} \\[1mm]
{\em High-Tech Research Center, Saitama Institute of Technology} 
\footnote{e-mail:sawa@sit.ac.jp}\\
{\em Okabe-machi, Saitama 369-0293, Japan}\\[2mm]
{September 2003}\\[15mm]


\begin{abstract}
A nonlinear supersymmetric(NLSUSY)  Einstein-Hilbert(EH)-type new action for unity of nature  
is obtained by performing the Einstein gravity analogue geomtrical arguments in 
high symmetry spacetime inspired by NLSUSY. 
The new action is unstable and breaks down spontaneously into E-H action with matter in 
ordinary Riemann spacetime. 
All elementary particles  except graviton are composed of the fundamental fermion "superon" of 
Nambu-Goldstone(NG) fermion of NLSUSY and regarded as the eigenstates of SO(10) super-Poincar\'e 
(SP) algebra, called superon-graviton model(SGM) of nature. 
Some phenomenological implications for the low energy particle physics and the cosmology are discussed.    
The linearization of NLSUSY including N=1 SGM action  is attempted explicitly 
to obtain the linear SUSY  local field theory, which is equivalent and renormalizable . 
%
%

%
\end{abstract}
\end{center}

\newpage

\section{Introduction}
The standard model(SM) is established  
as a unified model for the strong-electroweak interactions. 
Nevertheless, there still remains many unsolved problems, e.g. 
it can not explain the particle quantum numbers $(Q_{e},I,Y,color)$, i.e. \ 
${\bf 1 \times 2 \times 3}$ \ gauge \ structure, the three-generations structure of quarks and leptons 
and contains more than 28 arbitrary parameters(in the case of neutrino oscillations) 
even disregarding the mass generation mechanism for neutrino. 
The simple and beautiful extension to SU(5) GUT has serious 
difficulties, e.g. the life time of proton,$\cdots$, etc and is excluded so far. 
The SM and GUT equiptted naively with SUSY\cite{wz}\cite{va}\cite{gl}   
have improved the situations, e.g. the unification 
of the gauge couplings at about $10^{17}$, relatively stable proton(now threatened by experiments),etc., 
but they posess more arbitrary parameters which diminish the naturalness of the unification. 
Also SUSY model usually requires two times more number of elementary particles than non-SUSY model, 
e.g., at least about  60 for  SUSY SM  and 160 for  SUSY GUT in curved spacetime.   \\
SUSY is an essential notion to unify various topological and non-topological charges and 
gives a natural framework to unify spacetime and matter leading to the birth of supergravity theory(SUGRA)\cite{fnf}\cite{dz}. 
Unfortunately the maximally extended SO(8) SUGRA\cite{dn} is too small to accommodate 
all observed particles\cite{g}. 
The straightforward extension to SO(N) SUGRA with ${N>9}$ has a difficulty in writing down the action 
due to so called the no-go theorem on the gravitational interaction of massless elementary 
high spin$(>2)$ (gauge) field\cite{cm}\cite{hls}. (The massive high-spin field is another.)   \\
We think that from the viewpoints of simplicity and beauty of nature 
it is natural to attempt the accommodation of all observed  particles in a single
irreducible representation of a certain algebra(group) especially in the case of 
spacetime having a certain boundary(,i.e. a boundary condition).   
And the dynamics  is to be  described by the spontaneous breakdown of the high symmetry of spacetime by itself, 
which is encoded in (the nonliner realization of) the geometrical arguments of  spacetime. 
Facing so many fundamental elementary particles and so many arbitrary parameters, 
we are tempted to imagine that they are the composite objects and/or that they should be attributed to 
the particular topological and geometrical structure of spacetime. 
Also the no-go theorem does not exclude the possibility that the fundamental action, 
if it exists, posesses the high-spin degrees of freedom(d.o.f.) 
not as the elementary massless fields but as certain {\it composite \ eigenstates} 
of a certain symmetry (algebra)  of the fundamental action recasted by  the asymptotic local fields. 
Note that the relativistic hydrogen atom is solved by O(4) symmetry and that the particular structure 
(and the boundary) of the bulk materials induces the high-$T_{c}$ superconductivity containing 
s- and d-wave pairs. 
In this talk we would like to present a model along this scenario according to the chronological order 
of the studies.       \par 
The structure of this paper is as follows. 
In section 2, we show by the group theoretical arguments that three generations of quarks and leptons 
can be accomodated in the single irreducible representation of SO(10) SP algebra  
and propose SGM for a composite model of observed particles except graviton. 
In section 3, the fundamental SGM action of the vacuum EH-type is obtained.  
In section 4, the linearization of NLSUSY is discussed 
to obtain the the equivalent (broken) LSUSY theory, which is renormalizable. 
In section 5, SGM with spin 3/2 NG fermion is presented.  
In section 6, the cosmological meaning of SGM is discussed qualitatively. 
In section 7, as a summary some characteristic properties of SGM including many open questions 
are discussed briefly. 
\section{SGM and SO(10) SP Algebra}   
In this section we study a single irreducible representation which contains all observed 
elementary particles. 
By considering the structure of the helicity states of 
the representation of SP algebra it is natural to consider SO(N) extention of SUSY. 
We have shown that among all single irreducible representations of all SO(N) extended SP symmetries, 
the massless irreducible representations of SO(10) SP algebra(SPA) is {\it the only one} that accommodates 
${\it minimally}$ all observed particles including the graviton\cite{ks1}\cite{ks2}.  
That is, SO(10) SP is unique among all SO(N) SP extention of the franework of 
SGM discussed below\cite{ks1}.   \\
SO(10) SP contains 10 generators  $Q^{N}(N=1,2,..,10)$, which  are the ten dimensional
fundamental represemtations of SO(10) internal symmetry. 
We have decomposed  10 generators  $Q^{N}(N=1,2,..,10)$ into 
$\underline{10} =  \underline 5+ \underline 5^{*}$  with respect to SU(5) following 
$SO(10) \supset SU(5) \times U(1) $. 
For the  massless  case the little algebra of SO(10) SPA for the
supercharges in  the light-cone frame $P_{\mu}=\epsilon(1,0,0,1)$  becomes
after a suitable rescaling
\begin{equation}
\{ Q_{\alpha}^{M}, Q_{\beta}^{N} \}
=\{ \bar{Q}_{\dot\alpha}^{M}, \bar{Q}_{\dot\beta}^{N} \}=0, \quad
\{Q_{\alpha}^{M},\bar{Q}_{\dot\beta}^{N}\}
={\delta}_{{\alpha}1}{\delta}_{{\dot\beta}{\dot1}}{\delta}^{MN},
\label{algebra}
\end{equation}
where $\alpha,\beta=1,2$ and $M,N=1,2,...5$\cite{wb}\cite{ss}\cite{fwz}. 
By identifying the graviton with the Clifford vacuum $\mid\Omega\rangle$
(SO(10) singlet) satisfying
$Q_{\alpha}^{M} \mid\Omega\rangle=0$ 
and performing the ordinary procedures\cite{n}\cite{fsz} we obtain  
$2\cdot2^{10}$ dimensional irreducible representation 
of the little algebra (\ref{algebra}) of SO(10) SPA as follows\cite{ks1}:     
$\Bigl[\underline{1}(+2), \underline{10}(+{3 \over 2}),
\underline{45}(+1), \underline{120}(+{1 \over 2}),    \\ 
\underline{210}(0),     
\underline{252}(-{1 \over 2}),  
\underline{210}(-1),
\underline{120}(-{3 \over 2}),
\underline{45}(-2), 
\underline{10}(-{5 \over 2}), \underline{1}(-3)\Bigr]
+ \Bigl[ \mbox{CPT-conjugate} \Bigr]$,                          \\
where $\underline{d}(\lambda)$  represents
SO(10) dimension $\underline{d}$ and the helicity $\lambda$.                   
By noting that the helicities of these  states are
automatically determined by SO(10) SPA in the light-cone and that
$Q_{1}^{M}$ and  $\bar{Q}_{\dot{1}}^{M}$
satisfy the algebra of the annihilation and the creation operators for the
massless spin ${1 \over 2}$ particle, we speculate boldly that
these massless states spanned upon the  Clifford vacuum  $\mid\Omega(\pm2)\rangle$
are the massless (gravitational) eigenstates of spacetime and matter 
with SO(10) SP symmetric structure, which are composed of the
fundamental massless object $Q^{N}$, ${superon}$ with spin ${1 \over 2}$\cite{ks2}\cite{ks3}.
Because they correspond merely to all possible nontrivial combinations of the multiplications of
the spinor charges(i.e. {\it creation or annihilation operators}) of SO(10) SP algebra.
Simultaneously we can escape from the no-go theorem in a sence that 
we can write down the fundamental action with $N>9$. 
Therefore we regard $\underline 5+ \underline 5^{*}$  as $\it{a}$ ${superon}$-${quintet}$
and $\it{an}$ ${antisuperon}$-${quintet}$ . 
And we call the model containing the above towers of the helicity states are superon-graviton model(SGM).  
The justification of this  bold assumption is given later. 
Interestingly the composite model of matter based upon VA model\cite{va} was attemted 
long time ago\cite{bv}.    \\
To survey the physical implications of superon model for matter we assign the 
following SM quantum numbers to superons and adopt the following symbols. 
\begin{eqnarray}\underline{10} & = & \underline 5
+ \underline 5^{*}  \nonumber \\
& = & \Bigr[ Q_{a}(a=1,2,3) ,Q_{m}(m=4,5)\Bigl] + \Bigr[Q_{a}^{*}(a=1,2,3),Q_{m}^{*}(m=4,5)\Bigl], \nonumber  \\
& = & [(\underline 3, \underline 1;-{1 \over 3},-{1 \over 3},-{1 \over 3}), 
 (\underline 1, \underline 2;1, 0)] +
[(\underline 3^{*}, \underline 1;{1 \over 3},{1 \over 3},{1 \over 3}), 
 (\underline 1, \underline 2^{*};-1,0)],
\end{eqnarray}
where we have specified (${\underline{SU(3)},\underline{SU(2)}}$; electric  charges ) and 
${a=1,2,3}$ and ${m=4,5}$ represent the color and electroweak components of
superons respectively. 
Our model needs only five superons as the fiundamental elementary objects,  
which have suprisingly the same quantum numbers as the fundamental matter multiplet ${\underline 5}$ of 
SU(5) GUT\cite{gg} 
and satisfy the Gell-Mann--Nishijima relation.     \\
\begin{equation}
Q_{e}=I_{z} + {1 \over 2}(B-L).
\end{equation}
Consequently all ${2 \cdot 2^{10}}$  states are specified uniquely with respect to 
($SU(3), SU(2)$; electric charges)\cite{ks2}. \\
Here we suppose drastically that the fundamental action of SGM exists and that 
all such composite states are represented by local fields 
in a certain energy scale 
with $SU(3) \times SU(2) \times U(1)$ invariance, 
where by absorbing the lower helicity states through the superHiggs mechanism 
and through the diagonlizations of the mass terms  the high-spin fields become massive 
through the spontaneous symmetry breaking 
[SO(10) SPA upon the Clifford vacuum]
$\rightarrow$ [ \ $SU(3) \times SU(2) \times U(1)$ ] 
$\rightarrow$ [ \ $SU(3) \times U(1)$ ]. 
We have carried out the recombinations of the helicity states and 
found surprisingly that besides the gauge bosons for ${3 \times 2 \times 1}$ 
all the massless states necessary  for 
the SM with three generations of quarks and leotons appear in the surviving 
massless states (therefore, no sterile neutrinos)\cite{ks1}.      \\
As for the assignments of observed particles, we take for simplicity   
the following left-right symmetric assignment for quarks and  leptons 
by using the conjugate representations  naively , i.e. 
$( \nu_{l}, \it{l}^{-} )_{R}= (\bar\nu_{l}, \it{l}^{+} )_{L}$, etc\cite{ks2}. 
Furthermore as for the identification of the generation of quarks and leptons  
we assume simply that the states with more (color-) superons turn to acquiring larger masses 
in the low energy and no a priori mixings among genarations.
The surviving massless states identified with  SM(GUT) are  as follows.                       \\
(In the paper \cite{ks2} the right-handed neutrinos are denoted as a new particle, 
for the right-haded neutrinos were not observed at that time.)  \\
For three generations of leptons
[$({\nu}_{e}, e)$,  $({\nu}_{\mu}, \mu)$,  $({\nu}_{\tau}, \tau)$], we take
\begin{equation}
\Bigr[(Q_{m}{\varepsilon}_{ln}Q_{l}^{*}Q_{n}^{*}),
(Q_{m}{\varepsilon}_{ln}Q_{l}^{*}Q_{n}^{*}Q_{a}Q_{a}^{*}),
(Q_{a}Q_{a}^{*}Q_{b}Q_{b}^{*}Q_{m}^{*})\Bigl]
\end{equation}
\noindent and the conjugate states respectively. 
SGM contains initially four lepton generations and one of them 
(in the present case, ${Q_{m}Q_{a}Q_{a}^{*}}$) disappears by the superHiggs mechanism.  \\
For three generations of quarks [$( u, d )$, $( c, s )$, $( t, b )$],
we have ${\em uniquely}$
\begin{equation}
\Bigr[({\varepsilon}_{abc}Q_{b}^{*}Q_{c}^{*}Q_{m}^{*}),
({\varepsilon}_{abc}Q_{b}^{*}Q_{c}^{*}Q_{l}
{\varepsilon}_{mn}Q_{m}^{*}Q_{n}^{*}),
({\varepsilon}_{abc}Q_{a}^{*}Q_{b}^{*}Q_{c}^{*}Q_{d}Q_{m}^{*})\Bigl]
\end{equation}
\noindent and conjugate states respectively.   \\
For $SU(2) \times U(1)$ gauge bosons [ ${{W}^{+},\ Z,\ \gamma,\ {W}^{-}}$],
$SU(3)$ gluons [${G^{a}(a=1,2,..,8)}$], 
[$SU(2)$ Higgs Boson], [$( X,Y )$] leptoquark bosons in GUTs, 
and  a color- and $SU(2)$-singlet neutral gauge boson from
${{\underline 3} \times {\underline 3^{*}}}$ (which we call simply $S$  boson
to represent the singlet) we have   \\
${[Q_{4}Q_{5}^{*}, {1 \over \sqrt{2}}( Q_{4}Q_{4}^{*} \pm Q_{5}Q_{5}^{*}),
Q_{5}Q_{4}^{*}]}$, \\
${[Q_{1}Q_{3}^{*},Q_{2}Q_{3}^{*},-Q_{1}Q_{2}^{*},
{1 \over \sqrt{2}}(Q_{1}Q_{1}^{*}-Q_{2}Q_{2}^{*}), Q_{2}Q_{1}^{*},  
{1 \over \sqrt{6}}(2Q_{3}Q_{3}^{*}-Q_{2}Q_{2}^{*}-Q_{1}Q_{1}^{*})}$,  \\
${-Q_{3}Q_{2}^{*},Q_{3}Q_{1}^{*}]}$,  ${[{\varepsilon}_{abc}Q_{a}Q_{b}Q_{c}Q_{m}]}$, 
$[Q_{a}^{*}Q_{m}]$ and  ${Q_{a}Q_{a}^{*}}$, (and their conjugates) respectively.   \par 
Now in order to test the superon picture and to see the potential of superon-quintet model(SQM) 
of matter in the low energy we try to interpret the Feynman diagrams of SM(GUT) in terms of 
the superon pictures of all particles in  SM(GUT).  
We replace a single line of the propagator of a particle 
in the Feynman diagrams of SM(GUT) by the multiple lines of superons  
constituting each particle under the following two  assumptoions at the vertex;   \\  
(i) the analogue of the OZI-rule of the quark model and  \\
(ii) the superon number consevation.    \\
The assumption (i) is natural, for  all particles are  assigned to each  state of a single irreducible 
representation of SO(10) SP algebra. Fig.1 shows the corresponding forbidden superon-line Feynmann diagram 
showing the transition between the different eigenstates ${a}$ and ${b}$ ${\em without \ the \ interaction}$. 
Fig.2 is the allowed diagram showing a interacting(decay) vertex ${a \rightarrow b + c}$.
(ii) is a superon number conservation and gives naturally a selection rule at the vertex 
as read in Fig.2. \\
\beginpicture 
	\setcoordinatesystem units <1mm,1mm>  
	\setplotarea  
		x from -20 to 100, y from -15 to 20
	
	\arrow <2mm> [0.2,0.6] from 0 9 to 10 9  
	\arrow <2mm> [0.2,0.6] from 10 9 to 50 9  
	\setlinear 
	\plot 50 9 60 9 / 
	\arrow <2mm> [0.2,0.6] from 0 0 to 10 0 
	\arrow <2mm> [0.2,0.6] from 10 0 to 50 0  
	\setlinear 
	\plot 50 0 60 0 / 
	\arrow <2mm> [0.2,0.6] from 20 6 to 50 6 
	\arrow <2mm> [0.2,0.6] from 20 3 to 50 3  
	\setlinear 
	\plot 50 6 60 6 /
	\plot 50 3 60 3 /
	\circulararc 180 degrees 
		from 20 6 center at 20 4.5 
	\setlinear 
	\plot 20 -10 40 19 / 
	\plot 20 19 40 -10 / 
	\circulararc -30 degrees 
		from -5 4.5 center at 5 4.5 
	\circulararc 30 degrees 
		from -5 4.5 center at 5 4.5 
	\circulararc 30 degrees 
		from 65 4.5 center at 55 4.5 
	\circulararc -30 degrees 
		from 65 4.5 center at 55 4.5 
	\put {$a$} <-10mm,0mm> at 0 4.5
	\put {$b$} <10mm,0mm> at 60 4.5
\endpicture
Fig.1

\vspace {1cm}

\beginpicture 
	\setcoordinatesystem units <1mm,1mm>  
	\setplotarea  
		x from -20 to 100, y from -15 to 20
	\arrow <2mm> [0.2,0.6] from 0 3 to 15 3  
	\setlinear 
	\plot 15 3 25 3 / 
	\arrow <2mm> [0.2,0.6] from 25 3 to 45 13  
	\setlinear 
	\plot 45 13 60 20.5 / 
	\arrow <2mm> [0.2,0.6] from 27 1 to 46 10.5  
	\setlinear 
	\plot 45 10 61 18 / 
	\circulararc 30 degrees 
		from 63 20.5 center at 57 17.5 
	\circulararc -30 degrees 
		from 63 20.5 center at 57 17.5 
	\arrow <2mm> [0.2,0.6] from 0 -3 to 15 -3  
	\setlinear 
	\plot 15 -3 25 -3 / 
	\arrow <2mm> [0.2,0.6] from 25 -3 to 45 -13 
	\setlinear 
	\plot 45 -13 60 -20.5 / 
	
	\arrow <2mm> [0.2,0.6] from 27 -1 to 46 -10.5 
	\setlinear
	\plot 45 -10 61 -18 / 
	\circulararc 30 degrees 
		from 63 -20.5 center at 57 -17.5 
	\circulararc -30 degrees 
		from 63 -20.5 center at 57 -17.5 
	\circulararc 120 degrees 
		from 27 1 center at 27.5 0 
	\circulararc -30 degrees 
		from -4 0 center at 4 0 
	\circulararc 30 degrees 
		from -4 0 center at 4 0 
	\put {$a$} <-10mm,0mm> at 0 0
	\put {$b$} <6mm,3mm> at 60 19
	\put {$c$} <6mm,-3mm> at 60 -19
\endpicture
Fig.2

\vspace {1cm}

%
We have studied whether  Feynmann diagrams of SM and (SUSY)GUT are 
reproduced ( i.e., SGM allowed ) or not ( i.e., SGM forbidden ) by the superon pictures 
under the asummptions (i) and (ii).  \\
As an example of the allowed diagram,  $\beta$ decay( Fig.3 ) is drawn in Fig.4 in terms of superons. 
%


\beginpicture 
	\setcoordinatesystem units <1mm,1mm> 
	\setplotarea  
		x from -20 to 100, y from -20 to 30
	
	\multiput{
		\arrow <2mm> [0.2,0.6] from 0 0 to 10 0 
		\arrow <2mm> [0.2,0.6] from 10 0 to 40 0 
		\setlinear 
			\plot 40 0 50 0 /
	} [r] at 0 0 *2 0 5 / 
	\put {$d$} <-3mm,0mm> at 0 0 
	\put {$u$} <-3mm,0mm> at 0 5 
	\put {$d$} <-3mm,0mm> at 0 10 
	\put {$u$} <3mm,0mm> at 50 0 
	\put {$u$} <3mm,0mm> at 50 5 
	\put {$d$} <3mm,0mm> at 50 10 
	\circulararc -30 degrees 
		from -10 5 center at 3 5 
	\circulararc 30 degrees 
		from -10 5 center at 3 5 
	\circulararc -30 degrees 
		from 60 5 center at 47 5 
	\circulararc 30 degrees 
		from 60 5 center at 47 5 
	\put {$n^0$} <-3mm,0mm> at -10 5
	\put {$p^+$} <3mm,0mm> at 60 5
	
	\startrotation by 0.707 -0.707 about 25 0 
	\multiput {
	\setquadratic
		\plot 0 0 1 1 2 0 3 -1 4 0 /
	} [rB] <-7mm,-18mm> at 25 0 *4 4 0 / 

	\stoprotation 
	\put {$W^-$} at 28 -10
	
	\arrow <2mm> [0.2,0.6] from 39.5 -14.4 to 47 -14.4 
	\setlinear 
			\plot 47 -14.4 50 -14.4 /
	\put {$e^-$} <3mm,0mm> at 50 -14.4
	
	\arrow <2mm> [0.2,0.6] from 39.5 -14.4 to 45.5 -17.4 
	\setlinear 
			\plot 45.5 -17.4 48.5 -18.9 /
	\put {$\bar{\nu}{}_e$} <3mm,-1mm> at 48.5 -18.9
	
	\setdots <1mm> 
	\setlinear
		\plot -2.5 3 53 3 /
		\plot 56 0 56 -20 /
		\plot -3 -3 50 -25 /
	\circulararc -90 degrees 
		from 53 3 center at 53 0
	\circulararc -110 degrees 
		from 56 -20 center at 51 -20
	\circulararc 170 degrees 
		from -2.5 3 center at -2.5 0
	\setsolid 
\endpicture
Fig.3

\vspace {1cm}


\beginpicture 
	\setcoordinatesystem units <1mm,1mm> 
	\setplotarea  
		x from -20 to 100, y from -20 to 30
	
	\multiput{
		\arrow <2mm> [0.2,0.6] from 0 0 to 15 0 
		\arrow <2mm> [0.2,0.6] from 15 0 to 45 0 
		\setlinear 
			\plot 45 0 60 0 /
	} [r] at 0 5 *1 0 5 / 
	\put {$a^*$} <-3mm,0mm> at 0 5 
	\put {$b^*$} <-3mm,0mm> at 0 10 
	\put {$a^*$} <3mm,0mm> at 60 5 
	\put {$b^*$} <3mm,0mm> at 60 10 
		\arrow <2mm> [0.2,0.6] from 0 0 to 15 0 
		\arrow <2mm> [0.2,0.6] from 29.5 0 to 45 0 
		\setlinear 
			\plot 15 0 25 0 /
			\plot 45 0 60 0 /
	\put {$4^*$} <-3mm,0mm> at 0 0 
	\put {$5^*$} <3mm,0mm> at 60 0 
	\circulararc -30 degrees 
		from -10 5 center at 3 5 
	\circulararc 30 degrees 
		from -10 5 center at 3 5 
	\circulararc -30 degrees 
		from 70 5 center at 57 5 
	\circulararc 30 degrees 
		from 70 5 center at 57 5 
	\put {$d$} <-3mm,0mm> at -10 5
	\put {$u$} <3mm,0mm> at 70 5
	\startrotation by 0.707 -0.707 about 25 0 
		\arrow <2mm> [0.2,0.6] from 25 0 to 35 0 
		\arrow <2mm> [0.2,0.6] from 35 0 to 65 0 
		\setlinear 
			\plot 60 0 70 0 /
		\arrow <2mm> [0.2,0.6] from 28 3 to 35 3 
		\arrow <2mm> [0.2,0.6] from 46.5 3 to 65 3 
		\setlinear 
			\plot 35 3 42 3 /
			\plot 60 3 70 3 /
		\arrow <2mm> [0.2,0.6] from 53.8 6 to 65 6 
		\setlinear 
			\plot 60 6 70 6 /
	\put {$4^*$} <1mm,-1mm> at 70 0
	\put {$4$} <1mm,-1mm> at 70 3
	\put {$5$} <1mm,-1mm> at 70 6
	\circulararc -30 degrees 
		from 75 3 center at 67 3 
	\circulararc 30 degrees 
		from 75 3 center at 67 3 
	\put {$\bar{\nu}_e$} <1.5mm,-1.5mm> at 75 3
	\stoprotation 
	\arrow <2mm> [0.2,0.6] from 39.2 -10 to 60 -10 
	\setlinear 
		\plot 55 -10 65 -10 /
	\arrow <2mm> [0.2,0.6] from 42.5 -13 to 60 -13 
	\setlinear 
		\plot 60 -13 65 -13 /
	\arrow <2mm> [0.2,0.6] from 49.5 -16 to 60 -16 
	\setlinear 
		\plot 55 -16 65 -16 /
	\put {$5^*$} [l] <1mm,0mm> at 65 -16
	\put {$4^*$} [l] <1mm,0mm> at 65 -13
	\put {$5$} [l] <1mm,0mm> at 65 -10
	\circulararc -30 degrees 
		from 71 -13 center at 63 -13 
	\circulararc 30 degrees 
		from 71 -13 center at 63 -13 
	\put {$e^-$} <3mm,0mm> at 71 -13
	\setdots <1mm> 
	\setlinear
		\plot -10 15 70 15 /
		\plot 80 5 80 -30 /
		\plot -14.5 -4 58 -43.5 /
	\circulararc 150 degrees 
		from -10 15 center at -10 5
	\circulararc -90 degrees 
		from 70 15 center at 70 5
	\circulararc -120 degrees 
		from 80 -30 center at 65 -30
	\setsolid 
	\put {$W^-$} at 25 -15
\endpicture
Fig.4

\vspace {1cm}


We find many remarkable results\cite{ks2}\cite{ks3}\cite{ks4} and show below only some of them.  \\  
In the SM( the low energy); 
the  naturalness of the mixing of $K^{0}$-$\overline{K^{0}}$, 
$D^{0}$-$\overline{D^{0}}$ and  $B^{0}$-$\overline{B^{0}}$(the difference of the mass eigenstates and 
the electroweak eigenstates), 
no CKM-like (but the different origins beyond SM) mixings among the lepton generations, 
$\pi^{0} \longrightarrow 2\gamma$ as an ordinary tree diagram of the dominating decay mode, 
no  $\mu \longrightarrow e + \gamma$ despite compositeness, $\cdots$, etc.   \\  
Beyond the SM; 
${\nu_{e} \longleftrightarrow \nu_{\mu} \longleftrightarrow \nu_{\tau}}$ transitions,  
the origins of the observed (strong) CP-violation and their qualitative differences among $K^{0}$, 
$D^{0}$ and $B^{0}$, 
the tiny values of the SM Yukawa couplings constants as effective couplings, $\cdots$, etc.  \\ 
In (SUSY)GUT;
no (Fig.6) dangerous tree diagrams inducing(Fig.5) proton decay 
(without introducing R-parity by hand for SUSY GUT), $\cdots$ etc..  \\
Fig.5 shows as an example the dangerous Feynmann diagram  of the main decay mode of proton 
$p \rightarrow \pi^{0} + e$ in GUT. 
Fig.6 shows that the gauge couipling vertex(the dotted circle of Fig.5) can not be  reproduced in terms of superons,which indicates proton does not decay $p \rightarrow \pi^{0} + e$  at the tree level.
Also proton is stable against the decay  $p \rightarrow  K^{+} + \bar \nu$ in SUSY GUT, 
for  the dangerous box-type Feynmann diagram of the decay  $p \rightarrow  K^{+} + \bar \nu$ in SUSY GUT 
can not be  reproduced in the superon picture. 
\beginpicture 
	\setcoordinatesystem units <1mm,1mm> 
	\setplotarea  
		x from -50 to 50, y from -20 to 20
	\startrotation by 0 1 about 0 0 
	\multiput {
	\setquadratic
		\plot 0 0 0.5 0.5 1 0 1.5 -0.5 2 0 /
	} [rB] at 0 0 *2 2 0 / 
	\stoprotation 
	%
	%
	\arrow <2mm> [0.2,0.6] from -20 0 to -10 0
	\arrow <2mm> [0.2,0.6] from -10 0 to 10 0
	\setlinear 
	\plot 10 0 20 0 /
	%
	%
	\arrow <2mm> [0.2,0.6] from -20 6 to -10 6
	\arrow <2mm> [0.2,0.6] from -10 6 to 10 6
	\setlinear 
	\plot 10 6 20 6 /
	%
	%
	\arrow <2mm> [0.2,0.6] from -20 12 to -10 12
	\arrow <2mm> [0.2,0.6] from -10 12 to 10 12
	\setlinear 
	\plot 10 12 20 12 /
	\put {$u$} <-3mm,0mm> at -20 12
	\put {$u$} <-3mm,0mm> at -20 6
	\put {$d$} <-3mm,0mm> at -20 0
	\put {$u$} [l] <2mm,0mm> at 20 12
	\put {$\bar{u}$} [l] <2mm,0mm> at 20 6
	\put {$e^+$} [l] <2mm,0mm> at 20 0
	\setdots <1mm> 
	\circulararc 360 degrees 
		from 0 10 center at 0 6
	\setsolid 
	\put {$\times$} <4mm,-4mm> at 0 6
	\put {$\Bigg($} at -30 6
	\circulararc -30 degrees 
		from -30 6 center at -15 6 
	\circulararc 30 degrees 
		from -30 6 center at -15 6 
	\put {$p^+$} <-3mm,0mm> at -30 6
	\circulararc -30 degrees 
		from 28 9 center at 20 9
	\circulararc 30 degrees 
		from 28 9 center at 20 9 
	\put {$\pi^0$} <3mm,0mm> at 28 9
\endpicture
Fig.5

%
\beginpicture 
	\setcoordinatesystem units <1mm,1mm>  
	\setplotarea  
		x from -50 to 50, y from -20 to 20
	%
	%
	\arrow <2mm> [0.2,0.6] from -30 4 to -20 4
	\arrow <2mm> [0.2,0.6] from 10 4 to 20 4
	\setlinear 
	\plot -20 4 -10 4 /
	\plot 20 4 30 4 /
	%
	%
	\arrow <2mm> [0.2,0.6] from -30 0 to -20 0
	\arrow <2mm> [0.2,0.6] from 10 0 to 20 0
	\setlinear 
	\plot -20 0 -10 0 /
	\plot 20 0 30 0 /
	%
	%
	\arrow <2mm> [0.2,0.6] from -30 -4 to -20 -4
	\arrow <2mm> [0.2,0.6] from 10 -4 to 20 -4
	\setlinear 
	\plot -20 -4 -10 -4 /
	\plot 20 -4 30 -4 /
	\put {$a^*$} <-3mm,0mm> at -30 4
	\put {$b^*$} <-3mm,0mm> at -30 0
	\put {$5^*$} <-3mm,0mm> at -30 -4
	\put {$a$} [l] <2mm,0mm> at 30 4
	\put {$b$} [l] <2mm,0mm> at 30 0
	\put {$5$} [l] <2mm,0mm> at 30 -4
	\setdots <1mm> 
	\circulararc 360 degrees 
		from 0 15 center at 0 0
	\setsolid 
	\put {$\times$} at 0 0
	\circulararc -20 degrees 
		from -38 0 center at -20 0 
	\circulararc 20 degrees 
		from -38 0 center at -20 0 
	\put {$u$} <-3mm,0mm> at -38 0
	\circulararc -20 degrees 
		from 38 0 center at 20 0 
	\circulararc 20 degrees 
		from 38 0 center at 20 0 
	\put {$\bar{u}$} <3mm,0mm> at 38 0
	%
	%
	\arrow <2mm> [0.2,0.6] from 2 -10 to 2 -20
	\arrow <2mm> [0.2,0.6] from -2 -10 to -2 -20
	\setlinear 
	\plot 2 -20 2 -25 /
	\plot -2 -20 -2 -25 /
	\put {$c^*$} <0mm,-3mm> at -2 -25
	\put {$4$} <0mm,-3mm> at 2 -25
	\circulararc -30 degrees 
		from 0 -32 center at 0 -22 
	\circulararc 30 degrees 
		from 0 -32  center at 0 -22 
	\put {$\times$} <0mm,-3mm> at 0 -32
\endpicture
Fig.6
\vspace{1cm}

Among predicted new particles in the low energy ( i.e., the states escape from being absorbed )\cite{ks2}; \\
one lepton-type  electroweak-doublet $( \nu_{\Gamma}, \Gamma^{-} )$ with spin ${3 \over 2}$ 
with the mass of the electroweak scale $( \leq Tev)$  and   
an electroweak-singlet and double-charge Dirac particle  $E^{++}$  
with spin ${1 \over 2}$ which should have large mass$( >Tev)$    \\ 
are color-singlet states for matter and can be produced directly.   \\ 
And the effects of S gauge boson may be observed in the coming (high energy) experiments, 
particularly in $B^{0}$ (and  $D^{0}$) decay and in  the various mixings of the electroweak eigenstates.   \par
Considering SUSY SM is usually equipted with $SU(3) \times SU(2) \times U(1) \times U(1)$ 
superon-quintet model(SQM)\cite{ks2} of matter may be the most economic gauge model 
containing the three generations of quarks and leptons.      \par 
Finally we just remark about the (superHiggs) mass generation mechanism assumed boldly above. 
This may be probable if the symmetry between the states containing 
up to 5 superons  and above 5 superons is broken spontaneously.
And/or it is probale if the symmetry between 
the massless states with spin $J$ and ${J-1}$ $(J \leq 3)$ is broken. 
One half of the helicity states (e.g. the superpartners) become massive and decouple, 
provided the mass is huge. 
\section{Fundamental  Action  of  SGM }  
In this section we justify the {\it superon}  hypothesis. 
The supercharges $Q$ of VA action of NLSUSY\cite{va} is computed by the supercurrents\cite{ks0}
\begin{equation}
J^{\mu}(x)={1 \over i}\sigma^{\mu}\psi(x)
-\kappa \{ \mbox{the higher order terms of $\kappa$, $\psi(x)$ and }\  \partial\psi(x) \}.
\end{equation}
(6) means the field-current identity between the elementary NG spinor field
$\psi(x)$ and the supercurrent, which justifies our bold assumption
that the generator(super \\ 
charge) $Q^{N} \sim \int J^{0}{^N}dx^{3}$ (N=1,2,..10) of SO(10) SPA in the light-cone frame represents
the fundamental massless  particle, $superon$ $with$ $spin$ ${1 \over 2}$.
Therefore superon is the NG fermion of NLSUSY and  the fundamental theory of SGM for spacetime and matter 
at(above) the Planck scale is SO(10) NLSUSY in the
curved spacetime(corresponding  to the Clifford vacuum $\mid\Omega(\pm 2)\rangle$).  \\
It is well known that it is impossible to write down the action of SUGRA with $N>8$
due to so called the no-go theorem on massless high spin($>2$) field based upon the S-matrix arguments.   \\ 
However we show that disregarding {\it a priori} S-matrix constraints at the begining   and 
giving weight to the geometrical arguments we can construct N=10 extended SUSY theory 
containing the garvitational interaction.  \par
Extending the geometrical arguments of Einstein general relativity theory(EGRT) on Riemann spacetime 
to new (SGM) spacetime where besides the Minkowski coordinate $x^{a}$ 
the coset space coordinates $\psi$  for SL(2C) of ${superGL(4,R) \over GL(4,R)}$ 
turning to the NG fermion degrees of freedom(d.o.f.) are attached at every  spacetime point,  
we obtain the following N=10 SGM action\cite{ks3}; 
\begin{equation}
L_{SGM}=-{c^{3} \over 16{\pi}G}\vert w \vert(\Omega + \Lambda ),
\label{SGM}
\end{equation}
\begin{equation}
\vert w \vert=det{w^{a}}_{\mu}=det({e^{a}}_{\mu}+ {t^{a}}_{\mu}),  \quad
{t^{a}}_{\mu}={i\kappa^{4}  \over 2}(\bar{\psi}^{j}\gamma^{a}
\partial_{\mu}{\psi}^{j}
- \partial_{\mu}{\bar{\psi}^{j}}\gamma^{a}{\psi}^{j}),
\label{w}
\end{equation} 
where $w^{a}{_\mu}$ and $e^{a}{_\mu}$ are the vierbeins of unified SGM spacetime and Riemann spacetime 
of EGRT respectively, $\psi^{j}$($j=1,2, ..,10$,) is NG fermion(superon) originating from 
the coset space coordinates of ${N=10 \ superGL(4,R) \over GL(4,R)}$, G is the gravitational constant, 
${\kappa^{4} = ({c^{3}\Lambda \over 16{\pi}G}})^{-1} $ is a fundamental volume of 
four dimensional spacetime of VA model\cite{va},  
and $\Lambda$ is a  ${small}$ cosmological constant related to the strength of 
the superon-vacuum coupling constant. 
Therefore SGM contains two mass scales,  ${1 \over {\sqrt G}}$(Planck scale) in the first term describing 
the curvature energy  and $\kappa \sim {\Lambda \over G}(O(1))$ in the second term describing the 
vacuum energy of SGM.
$\Omega$ is a new scalar curvature analogous to the Ricci scalar curvature $R$ of EGRT, 
whose explicit expression is obtained  by just replacing ${e^{a}}_{\mu}(x)$  
by ${w^{a}}_{\mu}(x)$ in Ricci scalar $R$\cite{st1}.    \\
These results can be understood intuitively by observing that 
\begin{equation}
{w^{a}}_{\mu}(x) ={e^{a}}_{\mu}(x)+ {t^{a}}_{\mu}(x), 
\label{w2}
\end{equation} 
inspired  by 
\begin{equation}
\omega^{a}=dx^{a} + {i\kappa^{4}  \over 2}(\bar{\psi}^{j}\gamma^{a}
d{\psi}^{j}
- d{\bar{\psi}^{j}}\gamma^{a}{\psi}^{j})
\sim {w^{a}}_{\mu}dx^{\mu},
\label{va-form}
\end{equation} 
where $\omega^{a}$ is the NLSUSY invariant differential forms of 
VA\cite{va}, is invertible, i.e.,
\begin{equation}
w^{\mu}{_a}= e^{\mu}{_a}- t{^{\mu}}_a + {t^{\mu}}_{\rho}{t^{\rho}}_a 
- t{^{\mu}}_{\sigma} t{^{\sigma}}_{\rho}   t{^{\rho}}_a 
+t{^{\mu}}_{\kappa} t{^{\kappa}}_{\sigma}t{^{\sigma}}_{\rho}t{^{\rho}}_a + \cdots, 
\label{w-inverse}
\end{equation} 
which terminates with $(t)^{10}$ and $s_{\mu\nu} \equiv w^{a}{_\mu}\eta_{ab}w^{b}{_\nu}$ and 
$s^{\mu \nu}(x) \equiv w^{\mu}{_{a}}(x) w^{{\nu}{a}}(x)$ 
are a unified vierbein and a unified metric tensor in SGM spacetime\cite{ks3}\cite{st1}. 
It is straightforward to show 
${w_{a}}^{\mu} w_{{\mu}{b}} = \eta_{ab}$,  $s_{\mu \nu}{w_{a}}^{\mu} {w_{b}}^{\mu}= \eta_{ab}$, ..etc. 
[As shown in (\ref{w}), throughout the paper the first and the second indices of $t$ represent those of the $\gamma$-matrix 
and the derivative, respectively.]  
It seems natural that the ordinary vierbein and the stress-enery-momentum tensor of superon 
contribute equally to the vierbein of the unified (SGM) spacetime.        \\
The SGM action  (\ref{SGM}) is invariant at least under the following symmetries\cite{st2};
global SO(10), ordinary local GL(4R),  
the following new NLSUSY transformation 
\begin{equation}
\delta^{NL} \psi^{i}(x) ={1 \over \kappa^{2}} \zeta^{i} + 
i \kappa^{2} (\bar{\zeta}^{j}{\gamma}^{\rho}\psi^{j}(x)) \partial_{\rho}\psi^{i}(x),
\quad
\delta^{NL} {e^{a}}_{\mu}(x) = i \kappa^{2} (\bar{\zeta}^{j}{\gamma}^{\rho}\psi^{j}(x))\partial_{[\rho} {e^{a}}_{\mu]}(x),
\label{newsusy}
\end{equation} 
where $\zeta^{i}, (i=1,..10)$ is a constant spinor and  $\partial_{[\rho} {e^{a}}_{\mu]}(x) = 
\partial_{\rho}{e^{a}}_{\mu}-\partial_{\mu}{e^{a}}_{\rho}$, \\
the following GL(4R) transformations due to (\ref{newsusy})  
\begin{equation}
\delta_{\zeta} {w^{a}}_{\mu} = \xi^{\nu} \partial_{\nu}{w^{a}}_{\mu} + \partial_{\mu} \xi^{\nu} {w^{a}}_{\nu}, 
\quad
\delta_{\zeta} s_{\mu\nu} = \xi^{\kappa} \partial_{\kappa}s_{\mu\nu} +  
\partial_{\mu} \xi^{\kappa} s_{\kappa\nu} 
+ \partial_{\nu} \xi^{\kappa} s_{\mu\kappa}, 
\label{newgl4r}
\end{equation} 
where  $\xi^{\rho}=i \kappa^{2} (\bar{\zeta}^{j}{\gamma}^{\rho}\psi^{j}(x))$, 
and the following local Lorentz transformation on $w{^a}_{\mu}$ 
\begin{equation}
\delta_L w{^a}_{\mu}
= \epsilon{^a}_b w{^b}_{\mu}
\label{Lrw}
\end{equation}
with the local  parameter
$\epsilon_{ab} = (1/2) \epsilon_{[ab]}(x)$    
or equivalently on  $\psi$ and $e{^a}_{\mu}$
\begin{equation}
\delta_L \psi(x) = - {i \over 2} \epsilon_{ab}
      \sigma^{ab} \psi,     \quad
\delta_L {e^{a}}_{\mu}(x) = \epsilon{^a}_b e{^b}_{\mu}
      + {\kappa^{4} \over 4} \varepsilon^{abcd}
      \bar{\psi}{^j} \gamma_5 \gamma_d \psi{^j}
      (\partial_{\mu} \epsilon_{bc}).
\label{newlorentz}
\end{equation}
The local Lorentz transformation forms a closed algebra, for example, on $e{^a}_{\mu}(x)$ 
\begin{equation}
[\delta_{L_{1}}, \delta_{L_{2}}] e{^a}_{\mu}
= \beta{^a}_b e{^b}_{\mu}
+ {\kappa^{4} \over 4} \varepsilon^{abcd} \bar{\psi}{^j}
\gamma_5 \gamma_d \psi{^j}
(\partial_{\mu} \beta_{bc}),
\label{comLr1/2}
\end{equation}
where $\beta_{ab}=-\beta_{ba}$ is defined by
$\beta_{ab} = \epsilon_{2ac}\epsilon{_1}{^c}_{b} -  \epsilon_{2bc}\epsilon{_1}{^c}_{a}$.
The commutators of two new NLSUSY transformations (\ref{newsusy})  on $\psi(x)$ and  ${e^{a}}_{\mu}(x)$ 
are GL(4R), i.e. new NLSUSY (\ref{newsusy}) is the square-root of GL(4R); 
\begin{equation}
[\delta_{\zeta_1}, \delta_{\zeta_2}] \psi
= \Xi^{\mu} \partial_{\mu} \psi,
\quad
[\delta_{\zeta_1}, \delta_{\zeta_2}] e{^a}_{\mu}
= \Xi^{\rho} \partial_{\rho} e{^a}_{\mu}
+ e{^a}_{\rho} \partial_{\mu} \Xi^{\rho},
\label{com1/2-e}
\end{equation}
where 
$\Xi^{\mu} = 2i\kappa (\bar{\zeta}_2 \gamma^{\mu} \zeta_1)
      - \xi_1^{\rho} \xi_2^{\sigma} e{_a}^{\mu}
      (\partial_{[\rho} e{^a}_{\sigma]})$.
They show the closure of the algebra. 
(The ordinary GL(4R) invariance is trivial by the construction.) 
SGM action (\ref{SGM}) is invariant at least under\cite{st2}
\begin{equation}
[{\rm global \ new \ NLSUSY}] \otimes [{\rm local\ GL(4,R)}] \otimes [{\rm local\ Lorentz}] 
\otimes [{\rm global\ SO(10)}],  \\
\label{sgmsymm}
\end{equation}
which is isomorphic to SO(10)SP  whose single irreducible representation gives 
the group theoretical description of SGM\cite{ks2}.   \par
Note that the no-go theorem\cite{cm}\cite{hls} is overcome in a sence that 
the massless N-extended theory with 
$N>8$ has been written down explicitly.
Here we just mention that the superon d.o.f. can  be gauged away  
neither by the ordinary GL(4R) transformations of $e{^a}_{\mu}(x)$ 
connecting $x^{\mu}$ and $x^{a}$ nor by the local spinor translation, 
e.g. ${\delta \psi(x)=\zeta(x)}$, 
$\delta e^{a}{_\mu}(x)=
-i \kappa^{2}(\bar\zeta(x)\gamma^{a} \partial_{\mu}\psi(x)+\bar\psi(x)\gamma^{a} \partial_{\mu}\zeta(x))$ 
which is nothing but a translation(redefifition) of {\em the spinor coordinate} 
in SGM spacetime. 
Therefore the action (\ref{SGM}) is a nontrivial generalization of the  EH action.   
Further details are read in Sec.7.  \\
Also it should be noticed that SGM action  (\ref{SGM}) posesses two types of flat space which are 
not equivalent, i.e. SGM-flat($w{^a}_{\mu}(x) \rightarrow {\delta}{^a}_{\mu}$)  and 
Riemann flat($e{^a}_{\mu}(x) \rightarrow {\delta}{^a}_{\mu}$). 
This structure plays impotant roles in the cosmology of SGM (\ref{SGM}) discussed in Sec 6.   
The linearization of such a theory with a high nonlinearity is an interesting  and 
inevitable to obtain an equivalent local field theory which is renormalizable  
and describes the observed low energy (SM) physics.  
We discuss these problems in the next section.              \par
\section{ Toward Linearization of SGM }     
\subsection{Linearization of N=1 NLSUSY in flat spacetime}
In advance of the linearization of the SGM we investigate the 
linearization of VA model\cite{va} in detail.    
The linearization of VA model has been investigated by many authors\cite{ik}\cite{r}\cite{uz}
and  proved that   N=1 VA model of NLSUSY is  equivalent to 
N=1 scalar supermultiplet action of LSUSY which is renormalizable.       
The general arguments on the constraints which gives the relations 
between the linear and the nonlinear realizations of global SUSY  
have been established\cite{ik}. 
Following the general arguments we  have shown  explicitly\cite{stt1} that the nonrenormalizable 
N=1 VA model is equivalent to a renormalizable  action of a U(1) gauge 
supermultiplet of the linear SUSY\cite{wz} with 
the Fayet-Iliopoulos(FI) $D$ term\cite{fi} indicating a spontaneous SUSY breaking\cite{stt1}. 
Remarkably we find that the magnitude of FI $D$ term(vacuum value) 
is determined uniqely to reproduce the correct sign of  VA action 
and that a U(1) gauge field  constructed explicitly in terms of NG fermion fields 
is an axial vector field for N=1.    \\
An $N=1$ U(1) gauge supermultiplet is given by a real superfield\cite{ss} \cite{fwz}
\begin{eqnarray}
V(x, \theta, \bar\theta) 
&=& C + i \theta \chi - i \bar\theta \bar\chi 
+ {1 \over 2} i \theta^2 (M+iN) 
- {1 \over 2} i \bar\theta^2 (M-iN) 
- \theta \sigma^m \bar\theta v_m        \nonumber  \\
& &+ i \theta^2 \bar\theta \left( \bar\lambda      
+ {1 \over 2} i \bar\sigma^m \partial_m \chi \right) 
- i \bar\theta^2 \theta \left( \lambda 
+ {1 \over 2} i \sigma^m \partial_m \bar\chi \right) 
 + {1 \over 2} \theta^2 \bar\theta^2 
\left( D + {1 \over 2} \Box C \right), 
\label{V}
\end{eqnarray}
where $C(x)$, $M(x)$, $N(x)$, $D(x)$ are real scalar fields, 
$\chi_\alpha(x)$, $\lambda_\alpha(x)$ and 
$\bar\chi_{\dot\alpha}(x)$, $\bar\lambda_{\dot\alpha}(x)$ are 
Weyl spinors and their complex conjugates, and $v_m(x)$ is 
a real vector field. 
We adopt the notations in ref.\ \cite{wb}. 
{}Following refs.\ \cite{ik},  we define the superfield 
$\tilde V(x, \theta, \bar\theta)$ by 
\begin{equation}
\tilde V(x, \theta, \bar\theta) = V(x', \theta', \bar\theta'), 
\label{tildev}
\end{equation}
\begin{equation}
x'^{\,m}  =  x^m + i \kappa \left( 
\zeta(x) \sigma^m \bar\theta 
- \theta \sigma^m \bar\zeta(x) \right), 
\theta'  =  \theta - \kappa \zeta(x), \qquad
\bar\theta' = \bar\theta - \kappa \bar\zeta(x). 
\label{cov}
\end{equation}
$\tilde V$ may be expanded as (\ref{V}) in component fields  
$ \{ \tilde\phi_i(x) \} =\{ \tilde C(x), \tilde\chi(x), \bar{\tilde\chi}(x), \cdots \}$, 
which can be expressed by $C, \chi, \bar\chi, \cdots$ and $\zeta$, $\bar\zeta$ 
by using the relation (\ref{tildev}). $\kappa$ is now defined with the dimension $(length)^{2}$. 
They have the supertransformations of the form 
\begin{equation}
\delta \tilde\phi_i = - i \kappa \left( \zeta \sigma^m \bar\epsilon 
- \epsilon \sigma^m \bar\zeta \right) \partial_m \tilde\phi_i. 
\end{equation}
Therefore, a condition $\tilde\phi_i(x) = {\rm constant}$ is 
invariant under supertransformations. 
As we are only interested in the sector 
which only depends on the NG fields, 
we eliminate other degrees of freedom than the NG fields 
by imposing  SUSY invariant constraints 
\begin{equation}
\tilde C = \tilde\chi = \tilde M = \tilde N = \tilde v_m 
= \tilde \lambda = 0, \qquad
\tilde D = {1 \over \kappa}. 
\label{constraints}
\end{equation}
Solving these constraints we find that the original component 
fields $C$, $\chi$, $\bar\chi$, $\cdots$ can be 
expressed by the NG fields $\zeta$, $\bar\zeta$.
Among them, the leading terms in the expansion of the fields 
$v_m$, $\lambda$, $\bar\lambda$ and $D$, which contain 
gauge invariant degrees of freedom, in $\kappa$ are 
\begin{equation}
v_m  =  \kappa \zeta \sigma_m \bar\zeta + \cdot, 
\lambda  =  i \zeta 
- {1 \over 2} \kappa^2 \zeta 
\left( \zeta \sigma^m \partial_m \bar\zeta 
- \partial_m \zeta \sigma^m \bar\zeta \right) 
+ \cdot, 
%
%
D =  {1 \over \kappa} 
+ i \kappa \left( \zeta \sigma^m \partial_m \bar\zeta 
- \partial_m \zeta \sigma^m \bar\zeta \right) + \cdot, 
\label{relation2}
\end{equation}
where $\cdot$ are higher order terms in $\kappa$. 
Our discussion so far does not depend on a particular form 
of the action. We now consider a free action of a U(1) gauge 
supermultiplet of LSUSY with a FI $D$ term. 
In component fields we have 
\begin{equation}
S = \int d^4x \left[ -{1 \over 4} v_{mn} v^{mn} 
- i \lambda \sigma^m \partial_m \bar\lambda 
+ {1 \over 2} D^2 - {1 \over \kappa} D \right]. 
\label{gaugeaction}
\end{equation}
The last term proportional to $\kappa^{-1}$ is the 
FI $D$ term.
The field equation for $D$ gives 
$D = {1 \over \kappa} \not= 0$ in accordance with 
eq.\ (\ref{constraints}), 
which indicates the spontaneous breakdown of supersymmetry. 
We substitute eq.\ (\ref{relation2}) into the action 
(\ref{gaugeaction}) and obtain an action for the NG fields 
$\zeta$, $\bar\zeta$ which is exactly N=1 VA action. 
\begin{equation}
S = -{1 \over 2\kappa^2} \int d^4x \, \det \left[ \delta_m^n 
+ i \kappa^2 \left( \zeta \sigma^n \partial_m \bar\zeta 
- \partial_m \zeta \sigma^n \bar\zeta \right) \right]. 
\end{equation}
For N=1, U(1) gauge field  becomes 
$v_m \sim \kappa \bar\zeta \gamma_m \gamma_5 \zeta + \cdots$ 
in the four-component spinor notation, which is unfortunately an {\em axial} vector 
and can not be identified with the observed vector gauge boson of SM. 
However these are very  suggestive and favourable to SGM and as shown in the next section 
the {\em vector} gauge field and ${SU(2)}$ gauge structure appear simultaneously 
in N=2.           \par
\subsection{Linearization of N=2 NLSUSY in flat spacetime}
Next we focus our attention to the $N = 2$ SUSY 
and discuss a connection between the VA model and an $N = 2$ 
vector supermultiplet \cite{f} of the linear SUSY in 
four-dimensional spacetime. In particular, we show that for the 
$N = 2$ theory a SUSY invariant relation between component fields of 
the vector supermultiplet and the NG fermion fields can be 
constructed by means of the method used in Ref.\ \cite{r} starting 
from an ansatz given below (Eq.\ (\ref{ansatz})). We also briefly 
discuss a relation of the actions for the two models. 
\par
Let us denote the component fields of an $N = 2$ U(1) gauge 
supermultiplet \cite{f}, which belong to representations of 
a rigid SU(2) \cite{sw}, as follows; 
namely, $\phi$ for a physical complex scalar field, 
$\lambda_R^i$ $(i = 1, 2)$ for two right-handed Weyl spinor fields 
and $A_a$ for a U(1) gauge field in addition to $D^I$ $(I = 1,2,3)$ 
for three auxiliary real scalar fields 
required from the mismatch of the off-shell degrees of freedom 
between bosonic and fermionic physical fields.
$\lambda_R^i$ and $D^I$ belong to representations {\bf 2} and {\bf 3} 
of SU(2) respectively while other fields are singlets. 
By the charge conjugation we define left-handed spinor fields as 
$\lambda_{Li} = C \bar\lambda_{Ri}^T$. 
We use the antisymmetric symbols $\epsilon^{ij}$ and 
$\epsilon_{ij}$ ($\epsilon^{12} = \epsilon_{21} = +1$) to raise 
and lower SU(2) indices as $\psi^i = \epsilon^{ij} \psi_j$, 
$\psi_i = \epsilon_{ij} \psi^j$. 
\par
The $N=2$ LSUSY transformations of these component fields 
generated by constant spinor parameters $\zeta_L^i$ are 
\begin{eqnarray}
\delta_Q \phi= - \sqrt{2} \bar\zeta_R \lambda_L, \nonumber  \\
\delta_Q \lambda_{Li} 
= - {1 \over 2} F_{ab} \gamma^{ab} \zeta_{Li} 
- \sqrt{2} i \gamma^{a}\partial_{a} \phi \zeta_{Ri} 
+ i ( \zeta_L \sigma^I )_i D^I, \nonumber  \\
\delta_Q A_a \ = \ - i \bar\zeta_L \gamma_a \lambda_L 
- i \bar\zeta_R \gamma_a \lambda_R, \nonumber  \\
\delta_Q D^I {} = {} \bar\zeta_L \sigma^I \gamma^{a}\partial_{a}\lambda_L 
+ \bar\zeta_R \sigma^I \gamma^{a}\partial_{a} \lambda_R, 
\label{lsusy}
\end{eqnarray}
where $\zeta_{Ri} = C \bar\zeta_{Li}^T$, 
$F_{ab} = \partial_a A_b - \partial_b A_a$, and $\sigma^I$ are 
the Pauli matrices. The contractions of SU(2) indices 
are defined as 
$\bar\zeta_R \lambda_L = \bar\zeta_{Ri} \lambda_L^i$, 
$\bar\zeta_R \sigma^I \lambda_L 
= \bar\zeta_{Ri} (\sigma^I)^i{}_j \lambda_L^j$, etc. 
These supertransformations satisfy a closed off-shell 
commutator algebra 
\begin{equation}
[ \delta_Q(\zeta_1), \delta_Q(\zeta_2)] 
= \delta_P(v) + \delta_g(\theta), 
\label{commutator}
\end{equation}
where $\delta_P(v)$ and $\delta_g(\theta)$ are a translation and 
a U(1) gauge transformation with parameters 
\begin{equation}
v^a {} ={} 2i ( \bar\zeta_{1L} \gamma^a \zeta_{2L} 
- \bar\zeta_{1R} \gamma^a \zeta_{2R} ), \nonumber
\theta {}={}  - v^a A_a + 2 \sqrt{2} \bar\zeta_{1L} \zeta_{2R} \phi 
- 2 \sqrt{2} \bar\zeta_{1R} \zeta_{2L} \phi^*. 
\label{u1}
\end{equation}
Only the gauge field $A_a$ transforms under the U(1) gauge 
transformation 
\begin{equation}
\delta_g(\theta) A_a = \partial_a \theta. 
\end{equation}
\par
Although our discussion on the relation between the linear and 
NLSUSY transformations does not depend on a form of 
the action, it is instructive to consider 
a free action which is invariant under Eq.\ (\ref{lsusy}) 
\begin{equation}
S_{\rm lin} = \int d^4 x \left[ \partial_a \phi \partial^a \phi^* 
- {1 \over 4} F^2_{ab} 
+ i \bar\lambda_R \!\!\not\!\partial \lambda_R 
+ {1 \over 2} (D^I)^2 - {1 \over \kappa} \xi^I D^I \right], 
\label{lact}
\end{equation}
where $\kappa$ is a constant whose dimension is $({\rm mass})^{-2}$ 
and $\xi^I$ are three arbitrary real parameters satisfying 
$(\xi^I)^2 = 1$. The last term proportional to $\kappa^{-1}$ is an 
analog of the FI $D$ term in the $N=1$ theories \cite{fi}. 
The field equations for the auxiliary fields 
give $D^I = \xi^I / \kappa$ indicating a spontaneous SUSY breaking. 
\par
On the other hand, in the $N = 2$ VA model \cite{bv} we have 
a NLSUSY transformation law of the NG fermion 
fields $\psi_L^i$ 
\begin{equation}
\delta_Q \psi_L^i = {1 \over \kappa} \zeta_L^i 
- i \kappa \left( \bar\zeta_L \gamma^a \psi_L 
- \bar\zeta_R \gamma^a \psi_R \right) 
\partial_a \psi_L^i, 
\label{nlsusy}
\end{equation}
where $\psi_{Ri} = C \bar\psi_{Li}^T$. This transformation 
satisfies off-shell the commutator algebra (\ref{commutator}) 
without the U(1) gauge transformation on the right-hand side. 
The VA action invariant under Eq.\ (\ref{nlsusy}) reads 
\begin{equation}
S_{\rm VA} = - {1 \over {2 \kappa^2}} \int d^4 x \; \det w, 
\label{vaact}
\end{equation}
where the $4 \times 4$ matrix $w$ is defined by 
\begin{equation}
w{^a}_b = \delta^a_b + \kappa^2 t{^a}_b, \qquad 
t{^a}_b = - i \bar\psi_L \gamma^a \partial_b \psi_L 
+ i \bar\psi_R \gamma^a \partial_b \psi_R. 
\end{equation}
The VA action (\ref{vaact}) is expanded in $\kappa$ as 
\begin{eqnarray}
S_{\rm VA} &=&
-{1 \over {2 \kappa^2}} \int d^4 x   
\left[ 1 + \kappa^2 t{^a}_a  
+ {1 \over 2} \kappa^4 (t{^a}_a t{^b}_b 
- t{^a}_b t{^b}_a) \right. \nonumber  \\
& &
\left. - {1 \over 6} \kappa^6 \epsilon_{abcd} \epsilon^{efgd} 
t{^a}_e t{^b}_f t{^c}_g 
- {1 \over 4!} \kappa^8 \epsilon_{abcd} \epsilon^{efgh} 
t{^a}_e t{^b}_f t{^c}_g t{^d}_h 
\right]. 
\label{vaactex}
\end{eqnarray}
We would like to obtain a SUSY invariant relation between the 
component fields of the $N = 2$ vector supermultiplet and 
the NG fermion fields $\psi^i$ at the leading orders of $\kappa$. 
It is useful to imagine a situation in which the linear SUSY 
is broken with the auxiliary fields having expectation values 
$D^I = \xi^I / \kappa$ as in the free theory (\ref{lact}). 
Then, we expect from the experience in the $N = 1$ cases 
\cite{ik}\cite{r}\cite{uz}\cite{stt1} and the transformation law of the spinor 
fields in Eq.\ (\ref{lsusy}) that the relation should have a form 
\begin{equation}
\lambda_{Li}  = i \xi^I (\psi_L \sigma^I)_i 
+ {\cal O}(\kappa^2), \nonumber  \\
{}D^I =  {1 \over \kappa} \xi^I + {\cal O}(\kappa), \nonumber
({\rm other\ fields}) =  {\cal O}(\kappa). 
\label{ansatz}
\end{equation}
Higher order terms are obtained such that the LSUSY transformations (\ref{lsusy}) 
are reproduced by the NLSUSY transformation of the NG fermion fields (\ref{nlsusy}). 
\par
After some calculations we obtain the relation between the fields 
in the linear theory and the NG fermion fields as 
\ba
\phi(\psi) \A = \A {1 \over \sqrt{2}} \, i \kappa \xi^I 
\bar\psi_R \sigma^I \psi_L 
- \sqrt{2} \kappa^3 \xi^I \bar\psi_L \gamma^a \psi_L 
\bar\psi_R \sigma^I \partial_a \psi_L \nonu
\A\A - {\sqrt{2} \over 3} \kappa^3 \xi^I \bar\psi_R \sigma^J \psi_L 
\bar\psi_R \sigma^J \sigma^I \!\!\not\!\partial \psi_R 
+ {\cal O}(\kappa^5), \nonu
\lambda_{Li}(\psi) \A = \A i \xi^I (\psi_L \sigma^I)_i 
+ \kappa^2 \xi^I \gamma^a \psi_{Ri} \bar\psi_R \sigma^I 
\partial_a \psi_L 
+ {1 \over 2} \kappa^2 \xi^I \gamma^{ab} \psi_{Li} 
\partial_a \left( \bar\psi_L \sigma^I \gamma_b \psi_L \right) \nonu
\A\A + {1 \over 2} \kappa^2 \xi^I ( \psi_L \sigma^J )_i 
\left( \bar\psi_L \sigma^J \sigma^I \!\!\not\!\partial \psi_L 
- \bar\psi_R \sigma^J \sigma^I \!\!\not\!\partial \psi_R \right) 
+ {\cal O}(\kappa^4), \nonu
A_a(\psi) \A = \A - {1 \over 2} \kappa \xi^I \left( 
\bar\psi_L \sigma^I \gamma_a \psi_L 
- \bar\psi_R \sigma^I \gamma_a \psi_R \right) \nonu
\A\A + {1 \over 4} i \kappa^3 \xi^I \biggl[ 
\bar\psi_L \sigma^J \psi_R \bar\psi_R \left( 2 \delta^{IJ} \delta_a^b 
- \sigma^J \sigma^I \gamma_a \gamma^b \right) \partial_b \psi_L \nonu
\A\A - {1 \over 4} \bar\psi_L \gamma^{cd} 
\psi_R \bar\psi_R \sigma^I \left( 2 \gamma_a \gamma_{cd} \gamma^b 
- \gamma^b \gamma_{cd} \gamma_a \right) \partial_b \psi_L 
+ (L \leftrightarrow R) \biggr] + {\cal O}(\kappa^5), \nonu
D^I(\psi) \A = \A {1 \over \kappa} \xi^I 
- i \kappa \xi^J \left( 
\bar\psi_L \sigma^I \sigma^J \!\!\not\!\partial \psi_L 
- \bar\psi_R \sigma^I \sigma^J \!\!\not\!\partial \psi_R \right) \nonu
\A\A + \kappa^3 \xi^J \biggl[ \bar\psi_L \sigma^I \psi_R \partial_a 
\bar\psi_R \sigma^J \partial^a \psi_L 
- \bar\psi_L \sigma^K \gamma^c \psi_L \biggl\{ 
i \epsilon^{IJK} \partial_c \bar\psi_L \!\!\not\!\partial \psi_L 
\nonu
\A\A - {1 \over 2} \partial_a \bar\psi_L 
\sigma^J \sigma^K \sigma^I \gamma_c \partial^a \psi_L 
+ {1 \over 4} \partial_a \bar\psi_L \sigma^J \sigma^I \sigma^K 
\gamma^a \gamma_c \!\!\not\!\partial \psi_L \biggr\} \nonu
\A\A - {1 \over 4} \bar\psi_L \sigma^K \psi_R \left\{ 
\partial_a \bar\psi_R \sigma^J \sigma^I \sigma^K \gamma^b \gamma^a 
\partial_b \psi_L 
- \bar\psi_R \left( 2 \delta^{IK} + \sigma^I \sigma^K \right) 
\sigma^J \Box \psi_L \right\} \nonu
\A\A + {1 \over 16} \bar\psi_L \gamma^{cd} \psi_R \left\{ 
\partial_a \bar\psi_R \sigma^J \sigma^I \gamma^b \gamma_{cd} 
\gamma^a \partial_b \psi_L 
+ \bar\psi_R \sigma^I \sigma^J \gamma^b \gamma_{cd} \gamma^a 
\partial_a \partial_b \psi_L \right\} \nonu
\A\A + (L \leftrightarrow R) \biggr] + {\cal O}(\kappa^5). 
\label{relation}
\ea
The transformation of the NG fermion fields (\ref{nlsusy}) 
reproduces the transformation of the linear theory (\ref{lsusy}) 
except that the transformation of the gauge field 
$A_a(\psi)$ contains an extra U(1) gauge transformation 
\begin{equation}
\delta_Q A_a(\psi) = - i \bar\zeta_L \gamma_a \lambda_L(\psi) 
- i \bar\zeta_R \gamma_a \lambda_R(\psi) + \partial_a X, 
\end{equation}
where 
\begin{equation}
X = {1 \over 2} i \kappa^2 \xi^I \bar\zeta_L \left( 
2 \delta^{IJ} - \sigma^{IJ} \right) \psi_R 
\bar\psi_R \sigma^J \psi_L + (L \leftrightarrow R). 
\end{equation}
The U(1) gauge transformation parameter $X$ satisfies 
\begin{equation}
\delta_Q(\zeta_1) X(\zeta_2) 
- \delta_Q(\zeta_2) X(\zeta_1) = - \theta, 
\end{equation}
where $\theta$ is defined in Eq.\ (\ref{u1}). 
Due to this extra term the commutator of two supertansformations 
on $A_a(\psi)$ does not contain the U(1) gauge transformation 
term in Eq.\ (\ref{commutator}). 
This should be the case since the commutator on $\psi$ does not 
contain the U(1) gauge transformation term. 
For gauge invariant quantities like $F_{ab}$ the transformations 
exactly coincide with those of the LSUSY. 
In principle we can continue to obtain higher order terms in the 
relation (\ref{relation}) following this approach. 
However, it will be more useful to use the $N=2$ superfield 
formalism \cite{GSW} as was done in Refs.\ \cite{ik,r,uz,stt1} 
for the $N=1$ theories. 
\par
We note that the leading terms of $A_a$ in Eq.\ (\ref{relation}) 
can be written as 
\begin{equation}
A_a = - \kappa \xi^1 \bar\chi \gamma_5 \gamma_a \varphi 
+ i \kappa \xi^2 \bar\chi \gamma_a \varphi 
- {1 \over 2} \kappa \xi^3 \left( \bar\chi \gamma_5 \gamma_a \chi 
- \bar\varphi \gamma_5 \gamma_a \varphi \right) 
+ {\cal O}(\kappa^3), 
\end{equation}
where we have defined Majorana spinor fields 
\begin{equation}
\chi = \psi_L^1 + \psi_{R1}, \qquad 
\varphi = \psi_L^2 + \psi_{R2}. 
\end{equation}
When $\xi^1 = \xi^3 = 0$, this shows the {\it vector} nature of the U(1) 
gauge field\cite{stt2} as we expected. 
\par
The relation (\ref{relation}) reduces to that of the $N = 1$ 
SUSY by imposing, e.g.  $\psi_L^2 = 0$. 
When $\xi^1 = 1$, $\xi^2 = \xi^3 = 0$, we find 
$\lambda_{L2} = 0$, $A_a = 0$, $D^3 = 0$ and that the relation 
between $(\phi, \lambda_{L1}, D^1, D^2)$ and $\psi_L^1$ 
becomes that of the $N=1$ scalar supermultiplet obtained in 
Ref.\ \cite{r}. 
When $\xi^1 = \xi^2 = 0$, $\xi^3 = 1$, on the other hand, 
we find $\lambda_{L1} = 0$, $\phi = 0$, $D^1 = D^2 = 0$ 
and that the relation between $(\lambda_{L2}, A_a, D^3)$ and 
$\psi_L^1$ becomes that of the $N=1$ (axial) vector 
supermultiplet obtained in Refs.\ \cite{ik,stt1}. 
\par
Our result (\ref{relation}) does not depend on a form of the 
action for the linear SUSY theory. 
We discuss here the relation between the free linear SUSY action 
$S_{\rm lin}$ in Eq.\ (\ref{lact}) and the VA action 
$S_{\rm VA}$ in Eq.\ (\ref{vaact}). 
It is expected that they coincide when Eq.\ (\ref{relation}) is 
substituted into the linear action (\ref{lact}) as in the 
$N=1$ case \cite{ik,r,uz,stt1}. 
We have explicitly shown that $S_{\rm lin}$ indeed coincides 
with the VA action $S_{\rm VA}$ up to and including 
$O(\kappa^0)$ in Eq.\ (\ref{vaactex})\cite{stt2}. 
\par
Now we summarize our results in this section.    \\ 
We have constructed the SUSY invariant relation between the component fields 
of the $N = 2$ vector supermultiplet and the NG fermion fields 
$\psi_L^i$ at the leading orders of $\kappa$. 
We have explicitly shown that the U(1) gauge field $A_a$ 
has the vector nature in terms of the  two NG fermion fields 
in contrast to the models with the $N = 1$ SUSY \cite{stt2}. 
The vector state with two NG fermion fields belongs to the adjoint 
representation of SGM scenario as expected. 
The relation (\ref{relation}) contains three arbitrary real 
parameters $\xi^I/\kappa$, which can be regarded as the vacuum 
expectation values of the auxiliary fields $D^I$. 
When we put $\psi_L^2 = 0$, the relation reduces to that of 
the $N = 1$ scalar supermultiplet or that of the $N = 1$ vector 
supermultiplet depending on the choice of the parameters $\xi^I$. 
We have also shown that the free action $S_{\rm lin}$ 
in Eq.\ (\ref{lact}) with the FI $D$ term 
reduces to the VA action $S_{\rm VA}$ in Eq.\ (\ref{vaact}) 
at least up to and including $O(\kappa^0)$. 
From these results  we anticipate 
the equivalence of the action of $N$-extended standard 
supermultiplets of LSUSY to the $N$-extended VA action 
of NLSUSY, which is favorable for the SGM scenario.   \\
It is interesting that SU(2) gauge structure of the SM is explained naturally
if the gauge bosons are the SUSY-composites of SGM-type.       \\
The derivation of the equivalent interacting LSYSY theory is yet to be studied. 
\subsection{Linearizing  SGM}
In this section we would like to  attempt\cite{sts1}\cite{sts2}\cite{sts3} 
the linearization of the new EH type action(N=1 SGM action) to obtain the equivalent 
LSUSY theory in the low energy. \\
Considering a phenomenological potential of SGM, though qualitative and group theoretical, 
discussed in \cite{ks1}\cite{ks2} based upon the composite picture of LSUSY representation 
and the recent interest in NLSUSY in superstring(membrane) world, 
the linearization of NLSUSY in curved spacetime may be of some general interest.  \par 
The linearization of SGM is physically interesting in general, 
even if it poduced  an  existing SUGRA-like theory, for the consequent broken LSUSY theory 
is shown to be equivalent and gives a new insight into the fundamental structure 
of  nature behind the low energy effective theory.      \par
For convenience we review N=1 SGM action briefly.
SGM action is given by\cite{ks3}; 
\begin{equation}
L_{SGM}=-{c^{3} \over 16{\pi}G}\vert w \vert(\Omega + \Lambda),
\label{SGM1}
\end{equation}
\begin{equation}
\vert w \vert = {\rm det}{w^{a}}_{\mu} 
= {\rm det}({e^{a}}_{\mu}+ {t^{a}}_{\mu}),  \quad
{t^{a}}_{\mu}={i \kappa^{4}  \over 2}(\bar{\psi}\gamma^{a}
\partial_{\mu}{\psi}
- \partial_{\mu}{\bar{\psi}}\gamma^{a}{\psi}), 
\label{w1}
\end{equation} 
%
%
%
%
%
which is invariant at least under the following symmetry\cite{st2};
ordinary GL(4R),  
the following new NL SUSY transformation; 
\begin{equation}
\delta^{NL} \psi(x) ={1 \over \kappa^{2}} \zeta + 
i \kappa^{2} (\bar{\zeta}{\gamma}^{\rho}\psi(x)) \partial_{\rho}\psi(x),
\quad
\delta^{NL} {e^{a}}_{\mu}(x) = i \kappa^{2} (\bar{\zeta}{\gamma}^{\rho}\psi(x))\partial_{[\rho} {e^{a}}_{\mu]}(x),
\label{newsusy1}
\end{equation} 
where $\zeta$ is a constant spinor and  $\partial_{[\rho} {e^{a}}_{\mu]}(x) = 
\partial_{\rho}{e^{a}}_{\mu}-\partial_{\mu}{e^{a}}_{\rho}$, \\
the following GL(4R) transformations due to (\ref{newsusy1});  
\begin{equation}
\delta_{\zeta} {w^{a}}_{\mu} = \xi^{\nu} \partial_{\nu}{w^{a}}_{\mu} + \partial_{\mu} \xi^{\nu} {w^{a}}_{\nu}, 
\quad
\delta_{\zeta} s_{\mu\nu} = \xi^{\kappa} \partial_{\kappa}s_{\mu\nu} +  
\partial_{\mu} \xi^{\kappa} s_{\kappa\nu} 
+ \partial_{\nu} \xi^{\kappa} s_{\mu\kappa}, 
\label{newgl4r1}
\end{equation} 
where  $\xi^{\rho}=i \kappa^{2} (\bar{\zeta}{\gamma}^{\rho}\psi(x))$, 
and the following local Lorentz transformation on $w{^a}_{\mu}$; 
\begin{equation}
\delta_L w{^a}_{\mu}
= \epsilon{^a}_b w{^b}_{\mu}
\label{Lrw1}
\end{equation}
with the local  parameter
$\epsilon_{ab} = (1/2) \epsilon_{[ab]}(x)$    
or equivalently on  $\psi$ and $e{^a}_{\mu}$
\begin{equation}
\delta_L \psi(x) = - {i \over 2} \epsilon_{ab}
      \sigma^{ab} \psi,     \quad
\delta_L {e^{a}}_{\mu}(x) = \epsilon{^a}_b e{^b}_{\mu}
      + {\kappa^{4} \over 4} \varepsilon^{abcd}
      \bar{\psi} \gamma_5 \gamma_d \psi
      (\partial_{\mu} \epsilon_{bc}).
\label{newlorentz1}
\end{equation}
The local Lorentz transformation forms a closed algebra, for example, on $e{^a}_{\mu}(x)$ 
\begin{equation}
[\delta_{L_{1}}, \delta_{L_{2}}] e{^a}_{\mu}
= \beta{^a}_b e{^b}_{\mu}
+ {\kappa^{4} \over 4} \varepsilon^{abcd} \bar{\psi}
\gamma_5 \gamma_d \psi
(\partial_{\mu} \beta_{bc}),
\label{comLr1/21}
\end{equation}
where $\beta_{ab}=-\beta_{ba}$ is defined by
$\beta_{ab} = \epsilon_{2ac}\epsilon{_1}{^c}_{b} -  \epsilon_{2bc}\epsilon{_1}{^c}_{a}$.
The commutators of two new NL SUSY transformations (\ref{newsusy1})  on $\psi(x)$ and  ${e^{a}}_{\mu}(x)$ 
are GL(4R), i.e. new NL SUSY (\ref{newsusy1}) is the square-root of GL(4R); 
\begin{equation}
[\delta_{\zeta_1}, \delta_{\zeta_2}] \psi
= \Xi^{\mu} \partial_{\mu} \psi,
\quad
[\delta_{\zeta_1}, \delta_{\zeta_2}] e{^a}_{\mu}
= \Xi^{\rho} \partial_{\rho} e{^a}_{\mu}
+ e{^a}_{\rho} \partial_{\mu} \Xi^{\rho},
\label{com1/2-e1}
\end{equation}
where 
$\Xi^{\mu} = 2i (\bar{\zeta}_2 \gamma^{\mu} \zeta_1)
      - \xi_1^{\rho} \xi_2^{\sigma} e{_a}^{\mu}
      (\partial_{[\rho} e{^a}_{\sigma]})$.
They show the closure of the algebra. 
SGM action (\ref{SGM1}) is invariant at least under\cite{st2}
\begin{equation}
[{\rm global \ new \ NLSUSY}] \otimes [{\rm local\ GL(4,R)}] \otimes [{\rm local\ Lorentz}],  \\
\label{sgmsymm1}
\end{equation}
which is isomorphic to SP whose single irreducible representation gives 
the group theoretical description of SGM\cite{ks2}.  \par
In the preceeding section  the linearization has been carried out by using the superfield formalism 
and/or by the heuristic and intuitive arguments 
on the relations between the component fields of LSUSY and NLSUSY. 
In either case it is crucial to discover  the SUSY invariant relations  which connect the supermultiplets 
of L and NL theories and reproduce the LSUSY transformations.   \\
In abovementioned cases of the global SUSY in flat spacetime  the SUSY invariant relations are 
obtained straightforwardly, for L and NL supermultiplets are well undestood and 
the algebraic structures are the same SP. 
(However as demonstrated the naive application of the superfield technique have produced 
the free theory of the linear supermultiplet.)     \par
The situation  is rather different in SGM, for (i) the supermultiplet structure of 
the linearized theory of SGM is unknown except it is expected to be a broken LSUSY SUGRA-like theory 
containing graviton and a (massive) spin 3/2 field as  dynamical d.o.f. and 
(ii) the algebraic structure (\ref{sgmsymm1}) is changed into  SP.       \\
And we should seek the linearization which produces the interacting theory of the linearized 
supermultilet and  unifies all charges.  \\
Therefore  by the heuristic arguments and referring to SUGRA  we discuss for the moment 
the linearization of N=1 SGM.          \par
At first, we assume  faithfully to SGM scenario that; \\
(i) the linearized theory should contain the  spontaneously broken ${\it global}$ (at least) LSUSY  \\
(ii) graviton is an elementary field(not composite of superons coresponding to the vacuum of the Clifford algebra) 
in both L and NL theories   \\
(iii) the NLSUSY supermultiplet of SGM ($e{^a}_{\mu}(x)$, $\psi(x)$)  should be connected 
to the composite supermultiplet 
(${\tilde e}{^a}_{\mu}(e(x), \psi(x))$, ${\tilde \lambda}_{\mu}(e(x), \psi(x))$)  
for elementary graviton field and a composite (massive) spin 3/2 field of the SUGRA-like linearized theory. \par
From these assumptions and following the arguments performed in the flat space cases  we require that 
the SUGRA gauge transformation with the global spinor parameter ${\zeta}$ 
should hold for the supermultiplet (${\tilde e}{^a}_{\mu}(e, \psi)$, ${\tilde \lambda_{\mu}(e, \psi)}$)  
of the (SUGRA-like) linearized theory, i.e.,  \\
\begin{equation}
\delta {\tilde e}{^a}_{\mu}(e, \psi)  
      = i\kappa \bar{ \zeta} \gamma^{a} {\tilde \lambda_{\mu}(e, \psi)}, 
\label{sugral-21}
\end{equation} 
\begin{equation}
\delta {\tilde \lambda}_{\mu}(e, \psi)  
      = {2 \over \kappa}D_{\mu}{ \zeta}  
      = -{i \over \kappa}{\tilde \omega}^{ab}{ _\mu}(e, \psi)\sigma_{ab}{ \zeta}, 
\label{sugral-3/21} 
\end{equation} 
where  $\sigma^{ab} = {i \over 4}[\gamma^a, \gamma^b]$, 
$D_{\mu}=\partial_{\mu}-{i \over 2}{\tilde \omega}^{ab}{ _\mu}(e, \psi)\sigma_{ab}$, ${ \zeta}$ is a 
global spinor parameter and the variations in the left-hand side are induced by NLSUSY (\ref{newsusy1}). \par
We put the following  SUSY invariant relations which connect $e^{a}{_\mu}$ to ${\tilde e}{^a}_{\mu}(e, \psi)$;
\begin{equation}
{\tilde e}{^a}_{\mu}(e, \psi) = { e}{^a}_{\mu}(x).     
\label{relation-21}
\end{equation} 
This relation (\ref{relation-21}) is the assumption (ii) and holds simply the metric conditions.
Consequently the following covariant relation  is obtained by substituting  (\ref{relation-21}) 
into (\ref{sugral-21}) and  computing the variations under (\ref{newsusy1})\cite{sts1}; 
\begin{equation}
{\tilde \lambda}_{\mu}(e, \psi)  
      = \kappa \gamma_{a} \gamma^{\rho} \psi(x) \partial_{[\rho} e{^a}_{\mu]}.    
\label{relation-3/21} 
\end{equation} 
(As discussed later these should may be considered as the leading order of the expansions in $\kappa$ of 
SUSY invariant relations. The expansions terminate with $(\psi)^{4}$.)
Now we see LSUSY transformation 
%
%
induced by (\ref{newsusy1}) on the (composite) supermultiplet 
(${\tilde e}{^a}_{\mu}(e, \psi)$, ${\tilde \lambda}_{\mu}(e, \psi$)).    \\
The LSUSY transformation  on $\tilde e{^a}_{\mu}$ becomes  as follows. 
The left-hand side of (\ref{sugral-21}) gives
\begin{equation}
\delta {\tilde  e}{^a}_{\mu}(e, \psi) = \delta^{NL} {e^{a}}_{\mu}(x) 
= i \kappa^{2} (\bar{\zeta}{\gamma}^{\rho}\psi(x))\partial_{[\rho} {e^{a}}_{\mu]}(x). 
\label{susysgm-21} 
\end{equation} 
While substituting (\ref{relation-3/21}) into the righ-hand side of  (\ref{sugral-21}) we obtain 
\begin{equation}
i \kappa^{2} (\bar{\zeta}{\gamma}^{\rho}\psi(x))\partial_{[\rho} {e^{a}}_{\mu]}(x) + \cdots(extra \ terms).
\label{susysgm-2-l} 
\end{equation} 
These results show that  (\ref{relation-21}) and (\ref{relation-3/21}) are not  SUSY invariant relations 
and reproduce (\ref{sugral-21}) with unwanted extra terms which should be identified with the auxirialy 
fields. 
The commutator of the two LSUSY transformations induces GL(4R) with the field dependent parameters as follows;
\begin{equation}
[\delta_{\zeta_1}, \delta_{\zeta_2}]{\tilde  e}{^a}_{\mu}(e, \psi) 
= \Xi^{\rho} \partial_{\rho} {\tilde  e}{^a}_{\mu}(e, \psi)    
+ {\tilde  e}{^a}_{\rho}(e, \psi)\partial_{\mu} \Xi^{\rho},
\label{susysgmcom-21}
\end{equation}
where 
$\Xi^{\mu} = 2i (\bar{\zeta}_2 \gamma^{\mu} \zeta_1)
      - \xi_1^{\rho} \xi_2^{\sigma} e{_a}^{\mu}
      (\partial_{[\rho} e{^a}_{\sigma]})$.

On  ${\tilde \lambda}_{\mu}(e, \psi)$, the left-hand side of (\ref{sugral-3/21}) becomes 
apparently rather complicated; 
\ba
\delta {\tilde \lambda}_{\mu}(e, \psi)  
\A = \A {\kappa } \delta( \gamma_{a} \gamma^{\rho} \psi(x) \partial_{[\rho} e{^a}_{\mu]}) \nonu
\A = \A {\kappa } \gamma_{a}[ \delta^{NL}\gamma^{\rho} \psi(x) \partial_{[\rho} e{^a}_{\mu]} + 
  \gamma^{\rho} \delta^{NL} \psi(x) \partial_{[\rho} e{^a}_{\mu]} +
  \gamma^{\rho} \psi(x) \partial_{[\rho} \delta^{NL} e{^a}_{\mu]}]. 
\label{susysgm-3/21} 
\ea
However the commutator of the two LSUSY transformations induces the similar GL(4,R);
\begin{equation}
[\delta_{\zeta_1}, \delta_{\zeta_2}]{\tilde \lambda}_{\mu}(e, \psi)  
= \Xi^{\rho} \partial_{\rho} {\tilde \lambda}_{\mu}(e, \psi)  
+ {\tilde \lambda}_{\rho}(e, \psi)\partial_{\mu} \Xi^{\rho}.  
\label{susysgmcom-3/21}
\end{equation}                  
These results indicate that it is necessary to generalize  (\ref{sugral-21}), (\ref{sugral-3/21}) and 
(\ref{relation-3/21})  for obtaining SUSY invariant relations and for the closure of the algebra.
%
%
%
%
%
%
Furthermore  due to the complicated expression of LSUSY (\ref{susysgm-3/21}) which makes the physical and 
mathematical structures are obscure, we can hardly guess a linearized invariant action 
which is equivalent to SGM.       \par
Now we generalize the linearization by considering the auxirialy fields 
such that  LSUSY transformation on the linearized fields 
induces SP transformation.   \par
%
%
%
%
%
By comparing (\ref{sugral-3/21}) with (\ref{susysgm-3/21}) we understand that 
the local Lorentz transformation  plays a crucial role. 
As for the local Lorentz transformation on the linearized asymptotic fields corresponding 
to the observed particles (in the low energy), 
it is natural to take (irrespective of (\ref{newlorentz1})) the following forms   \\ 
\begin{equation}
\delta_L \tilde \lambda_{\mu}(x) = - {i \over 2} { \epsilon}_{ab}
      \sigma^{ab} \tilde \lambda_{\mu}(x),     \quad
\delta_L \tilde { e^{a}}_{\mu}(x) = {\epsilon}{^a}_b \tilde e{^b}_{\mu}, 
\label{lorentz1}
\end{equation}
where $\epsilon_{ab} = (1/2)\epsilon_{[ab]}(x)$ is a local parameter.   
%
%
%
%
%
%
In SGM  the local Lorentz transformations  (\ref{Lrw1}) 
and (\ref{newlorentz1}), 
i.e. the local Lorentz invariant gravitational interaction of superon, 
are introduced   by the geomtrical arguments in SGM spacetime\cite{st2} following  EGRT.  
While in SUGRA theory the local Lorentz  transfomation  invariance  (\ref{lorentz1}) 
is  realized as usual by introducing  the Lorentz spin connection ${\omega^{ab}}{_\mu}$. 
And the L SUSY transformation is defined successfully by the (Lorentz) covariant 
derivative containing the spin connection $\tilde \omega^{ab}{_\mu}(e,\psi)$ as seen in (\ref{sugral-3/21}), 
which causes the super-Poincar\'e algebra on the commutator of SUSY and is convenient for 
constructing the invariant action. 
Therefore in the linearized (SUGRA-like) theory the local Lorentz  transformation  invariance is expected 
to be realized as usual by defining (\ref{lorentz1}) and introducing the Lorentz spin connection 
$\omega^{ab}{_\mu}$.
We investigate how the spin connection $\tilde\omega^{ab}{_\mu}(e,\psi)$ appears 
in the linearized (SUGRA-like) theory through  the linearization process. 
This is also crucial for constructing a nontrivial (interacting) linearized action 
which has manifest invariances.  \par  
We discuss the Lorentz covariance of the  transformation by comparing (\ref{susysgm-3/21}) with 
the right-hand side of (\ref{sugral-3/21}). 
The direct computation of (\ref{sugral-3/21}) by using SUSY invariant relations (\ref{relation-21}) 
and (\ref{relation-3/21}) under (\ref{newsusy1}) produces complicated redundant terms 
as read off from (\ref{susysgm-3/21}).
The local Lorentz invariance of the linearized theory may become ambiguous and lose the manifest invariance. \\
For a simple  restoration of the manifest local Lorentz invariance 
we survey the possibility that such redundant terms may be adjusted  by  the d.o.f of 
the auxiliary fields in the linearized supermultiplet.
As for the auxiliary fields it is necessary for the closure of the off-shell superalgebra 
to include the equal number of the fermionic and the bosonic d.o.f. in the linearized supermultiplet. 
As new NL SUSY is a global symmetry, ${\tilde \lambda}_{\mu}$ has 16 fermionic d.o.f.. 
Therefore at least 4 bosonic d.o.f. must be added to the off-shell SUGRA supermultiplet 
with 12 d.o.f.\cite{sw}\cite{fv} and a vector field  may be a simple candidate.  \par
However, counting the bosonic d.o.f. present in the redundant terms corresponding to 
$\tilde\omega^{ab}{_\mu}(e,\psi)$, 
we may need a bigger supermultiplet  e.g. $16 + 4 \cdot 16 = 80$  d.o.f., to carry out the linearization, 
in which case a rank-3 tensor $\phi_{\mu\nu\rho}$ 
and a rank-2 tensor-spinor $\lambda_{\mu\nu}$ may be candidates for the auxiliary fields.     \par
Now we consider  the  simple modification of SUGRA transformations(algebra) by adjusting 
the (composite) structure of the (auxiliary) fields.  
We take, in stead of (\ref{sugral-21}) and  (\ref{sugral-3/21}), 
\begin{equation}
\delta {\tilde e}{^a}_{\mu}(x) 
      = i\kappa \bar{\zeta} \gamma^{a} {{\tilde \lambda}_{\mu}(x)} + \bar{\zeta}{\tilde \Lambda}{^a}_{\mu}, 
\label{newsugral-21}
\end{equation} 
\begin{equation}
\delta {\tilde \lambda}_{\mu}(x) 
      = {2 \over \kappa}D_{\mu}\zeta + {\tilde \Phi}_{\mu}\zeta  
      = -{i \over \kappa}\tilde \omega^{ab}{_\mu}\sigma_{ab}\zeta + {\tilde \Phi}_{\mu}\zeta , 
\label{newsugral-3/21} 
\end{equation} 
where  $\tilde \Lambda{^a}_{\mu}$ and  $\tilde \Phi_{\mu}$ represent auxiliary fields 
which are functionals of $e^{a}{_\mu}$ and $\psi$. 
We need $\tilde \Lambda{^a}_{\mu}$ term in (\ref{newsugral-21}) to alter (\ref{susysgm-21}), 
(\ref{susysgmcom-21}), (\ref{susysgm-3/21}) and  (\ref{susysgmcom-3/21}) 
toward that of super-Poincar\'e algebra of SUGRA.  
We attempt the restoration of the manifest local Lorentz invariance order by order by adjusting 
$\tilde \Lambda{^a}_{\mu}$ and  $\tilde \Phi_{\mu}$. 
In fact, the Lorentz spin connection  ${\omega}^{ab}{_\mu}(e)$(i.e. the leading order terms of 
$\tilde\omega^{ab}{_\mu}(e, \psi)$) of (\ref{newsugral-3/21}) is reproduced by taking the following one  
\begin{equation}
\tilde \Lambda{^a}_{\mu} = {\kappa^{2} \over 4}[ ie_{b}{^\rho}\partial_{[\rho}e{^b}_{\mu]}\gamma{^a}\psi 
- \partial_{[\rho}e_{\mid b \mid \sigma]}e{^b}_{\mu}\gamma^{a}\sigma^{\rho\sigma}\psi ], 
\label{auxlambda-1}
\end{equation} 
where (\ref{susysgmcom-21}) holds.
Accordingly $\tilde \lambda_{\mu}(e,\psi)$ 
is determined up to the first order in $\psi$ as follows; 
\begin{equation}
\tilde \lambda_{\mu}(e,\psi) 
= { 1 \over 2i\kappa}( i\kappa^{2}\gamma_{a} \gamma^{\rho} \psi(x) \partial_{[\rho} e{^a}_{\mu]}
- \gamma_{a}\tilde \Lambda{^a}_{\mu} ) = -{i \kappa \over 2}\omega^{ab}{_\mu}(e)\sigma_{ab}\psi,  
\label{lambda-o11}
\end{equation} 
which indicates the minimal Lorentz covariant gravitational interaction of superon. 
Sustituting  (\ref{lambda-o11}) into (\ref{newsugral-3/21}) we obtain the following new LSUSY 
transformation of $\tilde \lambda_{\mu}$(after Fiertz transformations) 
%
\begin{eqnarray}
\delta {\tilde \lambda}_{\mu}(e,\psi) 
& = & -{i \kappa \over 2} \{ \delta^{NL}\omega^{ab}{_\mu}(e)\sigma_{ab}\psi + 
   {{\omega}_{\mu}}^{ab}(e)\sigma_{ab} \delta^{NL}\psi \}    \nonumber  \\
& = & -{i \over {2 \kappa}} \omega^{ab}{_\mu}(e)\sigma_{ab} \zeta + 
   {i \kappa \over 2} \{ \tilde \epsilon^{ab}(e,\psi)\sigma_{ab}\cdot{}\omega^{ab}{_\mu}(e)\sigma_{cd}\psi +
   \cdots \}.
\label{varlambda-o11} 
\end{eqnarray}
%
Remarkably the local Lorentz transformations  of  ${\tilde \lambda}_{\mu}(e,\psi)$ (,i.e. the second term)
with the field dependent antisymmetric parameters $\tilde \epsilon^{ab}(e, \psi)$ is induced 
in addition to the intended ordinary global SUSY transformation. 
This shows that (\ref{lambda-o11}) is the SUSY invariant relations for $\tilde \lambda_{\mu}(e,\psi)$, 
for the SUSY transformation of (\ref{lambda-o11}) gives the right hand side of (\ref{newsugral-3/21}) 
with the extra terms. 
Interestingly  the commutator of the two LSUSY transformations  on (\ref{lambda-o11}) induces 
GL(4R); 
\begin{equation}
[\delta_{\zeta_1}, \delta_{\zeta_2}]{\tilde \lambda}_{\mu}(e, \psi)  
= \Xi^{\rho} \partial_{\rho} {\tilde \lambda}_{\mu}(e, \psi)  
+ \partial_{\mu} \Xi^{\rho}{\tilde \lambda}_{\rho}(e, \psi),  
\label{varlambda-o1-comm1}
\end{equation}                   
where $\Xi^{\rho}$ is the same field dependent parameter as given in (\ref{susysgmcom-21}).
(\ref{susysgmcom-21}) and (\ref{varlambda-o1-comm1}) show the closure of the algebra 
on SP algebra provided that the SUSY invariant relations (\ref{relation-21}) and (\ref{lambda-o11}) 
are adopted. 
These phenomena coincide with SGM scenario\cite{ks1}\cite{ks2} from the algebraic point of view, 
i.e. they are the superon-graviton composite (eigenstates) corresponding to the  linear representations 
of SP algebra. 
As for the redundant higher order terms in (\ref{varlambda-o11})  
we can  adjust them by considering the modified spin connection $\tilde\omega^{ab}{_\mu}(e, \psi)$ 
particularly with the contorsion terms and by recasting them in terms of  (the auxiliary field d.o.f.) 
$\tilde \Phi_{\mu}(e,\psi)$. 
In fact,  we found that the following supermultiplet containing 
160 (= 80 bosonic + 80 fermionic) d.o.f. may be 
the supermultiplet of the SUGRA-like LSUSY theory which is equivalent to SGM, i.e.,      \\
for 80 bosonic d.o.f.
\ba 
\A \A
[ \ \tilde e{^a}_{\mu}(e,\psi), a_{\mu}(e,\psi), b_{\mu}(e,\psi), M(e,\psi), N(e,\psi),  \nonumber   \\  
\A \A 
A_{\mu}(e,\psi), B_{\mu}(e,\psi), A{^a}_{\mu}(e,\psi), B{^a}_{\mu}(e,\psi), A{^{[ab]}}_{\mu}(e,\psi) \ ]
\label{80bosons1}
\ea
and for 80 fermionic d.o.f.  \\
\ba 
[ \ \tilde{\lambda}{}_{\mu\alpha}(e,\psi), \  \tilde{\Lambda}{^a}_{\mu\alpha}(e,\psi) \ ],
\label{80fermions1}
\ea
where $\alpha=1,2,3,4$ are indices for  Majorana spinor. The gauge d.o.f. of 
the local GL(4R) and the local Lorentz of the vierbein are subtracted. 
Note that the second line of (\ref{80bosons1}) and the second term of (\ref{80fermions1}) are 
equivalent to  the off-shell (auxiliary) field with spin 3 and spin 5/2, respectively.    \\
The ${\it a \ priori}$ gauge invariance for  $\tilde \lambda_{\mu\alpha}(e,\psi)$ is 
not necessary for massive case, which can be anticipated by the spontaneous SUSY breaking. 
For it is natural to suppose that the equivalent linear theory 
may be a coupled system of graviton and massive spin 3/2  with the spontaneous global SUSY breaking, 
which may be an analogue obtained by the super-Higgs mechanism 
in the spontaneous local SUSY breaking of N=1 SUGRA\cite{dz1}.  
Although at the moment the arguments are independent of the form of the action.   \\
By continuing the heuristic arguments order by order referring to the familiar SUGRA supermultiplet 
we find the following SUSY invariant relations up to $O(\psi^{2})$: 
\ba 
\tilde e{^a}_{\mu}(e,\psi) 
\A=\A 
e{^a}_{\mu}, 
\label{compo-2} \\
%
\tilde \lambda_{\mu}(e,\psi) 
\A=\A 
-i\kappa(\sigma_{ab}\psi)\omega^{ab}{}_{\mu},    
\label{compo-3/2} \\
%
%
\tilde{\Lambda}{}^a{}_{\mu}(e,\psi)
\A=\A
\frac{\kappa^2}{2}\epsilon^{abcd}(\gamma_5\gamma_d\psi)\omega_{bc\mu}, 
\label{compo-5/2} \\
%
A_{\mu}(e,\psi)
\A=\A 
\frac{i\kappa^2}{4}
[(\bar{\psi}\gamma^{\rho}\partial_{\rho}\tilde{\lambda}_{\mu})
-(\bar{\psi}\gamma^{\rho}\tilde{\lambda}_a)\partial_{\mu}e^a{}_{\rho}
-(\bar{\tilde{\lambda}}_{\rho}\gamma^{\rho}\partial_{\mu}\psi)]  \nonumber \\
\A \A
+\frac{\kappa^3}{4}
[(\bar{\psi}\sigma^{a\rho}\gamma^{b}\partial_{\rho}\psi)
(\omega_{\mu ba}+\omega_{ab\mu}) 
+(\bar{\psi}\sigma^{ab}\gamma^{c}\partial_{\mu}\psi)\omega_{cab} ] \nonumber \\
\A \A
+\frac{\kappa^2}{8}(\bar{\tilde{\lambda}}_{\mu}\sigma_{ab}\gamma^{\rho}\psi)\omega^{ab}{}_{\rho},  
\label{compo-Amu} \\
%
B_{\mu}(e,\psi)
\A=\A
\frac{i\kappa^2}{4}
[-(\bar{\psi}\gamma_5\gamma^{\rho}\partial_{\rho}\tilde{\lambda}_{\mu})
+(\bar{\psi}\gamma_5\gamma^{\rho}\tilde{\lambda}_a)\partial_{\mu}e^a{}_{\rho}
-(\bar{\tilde{\lambda}}_{\rho}\gamma_5\gamma^{\rho}\partial_{\mu}\psi)]   \nonumber \\
\A \A
+\frac{\kappa^3}{4}
[(\bar{\psi}\gamma_5\sigma^{a\rho}\gamma^{b}\partial_{\rho}\psi)
(\omega_{\mu ba}+\omega_{ab\mu}) 
+(\gamma_5\sigma^{ab}\gamma^{c}\partial_{\mu}\psi)\omega_{cab} ] \nonumber \\
\A \A
+\frac{\kappa^2}{8}(\bar{\tilde{\lambda}}_{\mu}\gamma_5\sigma_{ab}\gamma^{\rho}\psi)\omega^{ab}{}_{\rho}, 
\label{compo-Bmu} \\
%
A^a{}_{\mu}(e,\psi)
\A=\A
\frac{i\kappa^2}{4}
[(\gamma^{\rho}\gamma^a\partial_{\rho}\tilde{\lambda}_{\mu})
-(\gamma^{\rho}\gamma^a\tilde{\lambda}_b)\partial_{\mu}\tilde{e}{}^b{}_{\rho}
+(\bar{\tilde{\lambda}}_{\rho}\gamma^a\gamma^{\rho}\partial_{\mu}\psi)]   \nonumber \\
\A \A
+\frac{\kappa^3}{4}
[-(\bar{\psi}\sigma^{b\rho}\gamma^a\gamma^{c}\partial_{\rho}\psi)
(\omega_{\mu cb}+\omega_{bc\mu})
-(\gamma^{bc}\sigma^a\gamma^{d}\partial_{\mu}\psi)\omega_{dbc} ]   \nonumber \\
\A \A
-\frac{\kappa^2}{8}(\bar{\tilde{\lambda}}_{\mu}\sigma_{bc}\gamma^a\gamma^{\rho}\psi)\omega^{ab}{}_{\rho}, 
\label{compo-Aamu} \\
%
%
B^a{}_{\mu}(e,\psi)
\A=\A 
\frac{i\kappa^2}{4}
[(\bar{\psi}\gamma_5\gamma^{\rho}\gamma^a\partial_{\rho}\tilde{\lambda}_{\mu})
-(\gamma_5\gamma^{\rho}\gamma^a\tilde{\lambda}_b)\partial_{\mu}\tilde{e}{}^b{}_{\rho}
+({\tilde{\lambda}}_{\rho}\gamma_5\gamma^a\gamma^{\rho}\partial_{\mu}\psi)]    \nonumber \\
\A \A
+\frac{\kappa^3}{8}
[-(\bar{\psi}\gamma_5\sigma^{b\rho}\gamma^a\gamma^{c}\partial_{\rho}\psi)
(\omega_{\mu cb}+\omega_{bc\mu})
-(\bar{\psi}\gamma_5\sigma^{bc}\gamma^a\gamma^{d}\partial_{\mu}\psi)\omega_{dbc} ]   \nonumber \\
\A \A
-\frac{\kappa^2}{8}(\bar{\tilde{\lambda}}_{\mu}\gamma_5\sigma_{bc}\gamma^a\gamma^{\rho}\psi)\omega^{ab}{}_{\rho},  
\label{compo-Bamu} \\
%
A^{[ab]}{}_{\mu}(e,\psi)
\A=\A
\frac{i\kappa^2}{2}
[(\bar{\psi}\gamma^{\rho}\sigma^{ab}\partial_{\rho}\tilde{\lambda}_{\mu})
-(\bar{\psi}\gamma^{\rho}\sigma^{ab}\tilde{\lambda}_c)\partial_{\mu}\tilde{e}{}^c{}_{\rho}
+(\bar{\tilde{\lambda}}_{\rho}\sigma^{ab}\gamma^{\rho}\partial_{\mu}\psi)]       \nonumber \\
\A \A
-\frac{\kappa^3}{2}
[(\bar{\psi}\sigma^{c\rho}\sigma^{ab}\gamma^{d}\partial_{\rho}\psi)
(\omega_{\mu dc}+\omega_{cd\mu})
+(\bar{\psi}\sigma^{cd}\sigma^{ab}\gamma^{e}\partial_{\mu}\psi)\omega_{ecd} ]   \nonumber \\
\A \A
-\frac{\kappa^2}{4}(\bar{\tilde{\lambda}}_{\mu}\sigma_{cd}\sigma^{ab}\gamma^{\rho}\psi)\omega^{ab}{}_{\rho}. 
\label{compo-Aabmu}
\ea
%
%
%
%
In fact we can show that the following LSUSY transformations on (\ref{80bosons1}) and (\ref{80fermions1}) 
inuced by  NLSUSY (\ref{newsusy1}) close among them(80+80 linearized multiplet). 
We show  the explicit expressions of some of the LSUSY transformations up to  $O(\psi)$. 
\ba
\delta \tilde{e}{}^a{}_{\mu} 
\A=\A
i\kappa \bar{\zeta}\gamma^a\tilde{\lambda}_{\mu}
-\epsilon^a{}_b\tilde{e}{}^b{}_{\mu}
+\bar{\zeta}\tilde{\Lambda}^a{}_{\mu} , 
\label{lsusy-compo-2} \\
\delta \tilde{\lambda}{}_{\mu} 
\A=\A 
-\frac{i}{\kappa}(\sigma_{ab}\zeta)\omega^{ab}{}_{\mu}
+\frac{i}{2}\epsilon^{ab}(\sigma_{ab}\tilde{\lambda}_{\mu})    \nonumber \\
\A \A
+A_{\mu}\zeta +B_{\mu}(\gamma_5\zeta)+A^a{}_{\mu}(\gamma_a\zeta)
+B^a{}_{\mu}(\gamma_5\gamma_a\zeta)+A^{ab}{}_{\mu}(\sigma_{ab}\zeta),
\label{lsusy-compo-3/2} \\
\delta \tilde{\Lambda}{^a}_{\mu}
\A=\A
\frac{1}{2}\epsilon^{abcd}(\gamma_5\gamma_d\zeta)\omega_{bc\mu},
\label{lsusy-compo-5/2} \\
%
\delta A{}_{\mu}
\A=\A
-\frac{1}{8}\left[
	i(\bar{\zeta}\gamma^{\rho}D_{\rho}\tilde{\lambda}_{a})\tilde{e}{}^a{}_{\mu}
	+3i(\bar{\zeta}\gamma^{a}D_{\mu}\tilde{\lambda}_{a})
	+2(\bar{\zeta}\sigma^{\nu\rho}\gamma_{\mu}D_{\nu}\tilde{\lambda}_{\rho})
	\right]
	\nonumber \\
\A \A
-\frac{1}{4\kappa}
	\left[
	3(\bar{\zeta}D_{\mu}\tilde{\Lambda}^a{}_{a})
	+i(\bar{\zeta}\sigma^{ab}D_{\mu }\tilde{\Lambda}_{ab})
	+i(\bar{\zeta}\sigma^{a\rho}D_{\rho }\tilde{\Lambda}_{(ab)})\tilde{e}{}^b{}_{\mu}
	\right]
	\nonumber \\
\A \A
+\frac{1}{16}
	\left[
	4i(\bar{\zeta}\gamma^{\rho}\tilde{\lambda}_a)\omega^a{}_{\rho\mu}
	+4(\bar{\zeta}\sigma^{bc}\gamma^a\tilde{\lambda}{}_{a})\omega_{bc\mu}
	-4(\bar{\zeta}\sigma^{a\rho}\gamma^b\tilde{\lambda}_{[\rho})\omega_{|ab|\mu ]}
	\right.
\nonumber \\
\A \A
	\hspace{1cm}\left.
	+4(\bar{\zeta}\sigma^{ab}\gamma^{c}\tilde{\lambda}_{a})\omega_{\mu cb}
	-3(\bar{\zeta}\sigma^{\rho}\gamma^{bc}\tilde{\lambda}_{[\rho})\omega_{|bc|\mu ]}
	+2i(\bar{\zeta}\sigma^{ab}\gamma_{\mu}\sigma^{cd}\tilde{\lambda}_{a})\omega_{cdb}
	\right]
	\nonumber \\
\A \A
-\frac{1}{8\kappa}
	\left[
	(\bar{\zeta}\gamma^{b}\gamma^a\sigma^{cd}{\tilde{\Lambda}}_{ab})
	+(\bar{\zeta}\sigma^{cd}\gamma^{b}\gamma^a{\tilde{\Lambda}}_{ab})
	\right]\omega_{cd\mu} , 
\label{lsusy-compo-1} \\
\delta A^a{}_{\mu}
\A=\A
\frac{1}{8}\left[
	-4i(\bar{\zeta}D_{\mu}\tilde{\lambda}^{a})
	+i(\bar{\zeta}\gamma^a\gamma^{\rho}D_{[\mu}\tilde{\lambda}_{\rho ]})
	+2(\bar{\zeta}\sigma^{\nu\rho}\gamma^a\gamma_{\mu}D_{\nu}\tilde{\lambda}_{\rho})
	\right]
	\nonumber \\
\A \A
+\frac{1}{4\kappa}
	\left[
	-i(\bar{\zeta}\sigma^{b\rho}\gamma^aD_{[\mu}\tilde{\Lambda}_{|b|\rho ]})
	-i(\bar{\zeta}\sigma^{\nu\rho}\gamma^aD_{\nu }\tilde{\Lambda}_{b\rho})\tilde{e}{}^b{}_{\mu}
	+(\bar{\zeta}\gamma^c\gamma^b\gamma^aD_{\mu}\tilde{\Lambda}_{bc})
	\right]
	\nonumber \\
\A \A
+\frac{1}{16}
	\left[
	-4i(\bar{\zeta}\gamma^{\rho}\gamma^a\tilde{\lambda}_b)\omega^b{}_{\rho\mu}
	-2(\bar{\zeta}\gamma^{\rho}\gamma^a\sigma^{bc}\tilde{\lambda}_{[\rho})\omega_{|bc|\mu ]}
	+2(\bar{\zeta}\gamma^a\sigma^{cd}\gamma^b\tilde{\lambda}_{b})\omega_{cd\mu}
	\right.
	\nonumber \\
\A \A
	\hspace{1cm}\left.
	+2(\bar{\zeta}\sigma^{cd}\gamma^a\gamma^b\tilde{\lambda}_{b})\omega_{cd\mu}
	+4(\bar{\zeta}\sigma^{b\rho}\gamma^a\gamma^c\tilde{\lambda}_{[\rho})\omega_{|bc|\mu ]}
	-4(\bar{\zeta}\sigma^{bc}\gamma^a\gamma^{d}\tilde{\lambda}_{b})\omega_{\mu dc}
	\right.
	\nonumber \\
\A \A
	\hspace{1cm}\left.
	-(\bar{\zeta}\gamma^a\gamma^{\rho}\sigma^{cd}\tilde{\lambda}_{[\rho})\omega_{|cd|\mu ]}
	-2(\bar{\zeta}\sigma^{bc}\gamma^a\gamma_{\mu}\sigma^{de}\tilde{\lambda}_{b})\omega_{dec}
	\right]
	\nonumber \\
\A \A
+\frac{1}{8\kappa}
	\left[
	(\bar{\zeta}\sigma^{b\rho}\gamma^a\sigma^{cd}\tilde{\Lambda}_{b[\rho})\omega_{|cd|\mu ]}
	-(\bar{\zeta}\sigma^{\nu\rho}\gamma^{a}\sigma^{bc}\tilde{\Lambda}_{\mu\nu})\omega_{bc\rho}
	+i(\bar{\zeta}\gamma^c\gamma^b\gamma^a\sigma^{de}\tilde{\Lambda}_{bc})\omega_{de\mu}
	\right.
	\nonumber \\
\A \A
	\hspace{1cm}\left.
	+i(\bar{\zeta}\sigma^{de}\gamma^c\gamma^b\gamma^a\tilde{\Lambda}_{bc})\omega_{de\mu}
	\right]
	\nonumber \\
\A \A
+\frac{\kappa}{2}(\bar{\zeta}D_{\mu}\Lambda^{\prime}{}^a) 
-\frac{\kappa}{4}(\bar{\zeta}\gamma^c\gamma^a\Lambda^{\prime}{}^b)\omega_{bc\mu}, 
\label{lsusy-compo-3} \\
\delta A^{[ab]}{}_{\mu}
\A=\A
\frac{1}{4}\left[
	-2i(\bar{\zeta}\gamma^{\rho}\sigma^{ab}D_{\rho}\tilde{\lambda}_{c})\tilde{e}{}^c{}_{\mu}
	+i(\bar{\zeta}\sigma^{ab}\gamma^{\rho}D_{\rho}\tilde{\lambda}_{c})\tilde{e}{}^c{}_{\mu}
	+i(\bar{\zeta}\sigma^{ab}\gamma^cD_{\mu}\tilde{\lambda}{}_{c}) 
	-2(\bar{\zeta}\sigma^{\nu\rho}\sigma^{ab}\gamma_{\mu}D_{\nu}\tilde{\lambda}_{\rho})
	\right]
	\nonumber \\
\A \A
+\frac{1}{2\kappa}
	\left[
	-(\bar{\zeta}\sigma^{ab}D_{\mu}\tilde{\Lambda}^{c}{}_{c}) 
	+i(\bar{\zeta}\sigma^{cd}\sigma^{ab}D_{\mu}\tilde{\Lambda}_{cd})
	+i(\bar{\zeta}\sigma^{c\rho}\sigma^{ab}D_{\rho }\tilde{\Lambda}_{(cd)})\tilde{e}{}^d{}_{\mu}
	\right]
	\nonumber \\
\A \A
+\frac{1}{8}
	\left[
	4i(\bar{\zeta}\gamma^{\rho}\sigma^{ab}\tilde{\lambda}_c)\omega^c{}_{\rho\mu}
	+4(\bar{\zeta}\sigma^{c\rho}\sigma^{ab}\gamma^d\tilde{\lambda}_{[\rho})\omega_{|cd|\mu ]}
	-4(\bar{\zeta}\sigma^{cd}\sigma^{ab}\gamma^{e}\tilde{\lambda}_{c})\omega_{\mu ed}
	\right.
\nonumber \\
\A \A
	\hspace{1cm}\left.
	-(\bar{\zeta}\sigma^{ab}\gamma^{\rho}\sigma^{de}\tilde{\lambda}_{[\rho})\omega_{|de|\mu ]}
	-2i(\bar{\zeta}\sigma^{cd}\sigma^{ab}\gamma_{\mu}\sigma^{ef}\tilde{\lambda}_{c})\omega_{efd}
	\right.
	\nonumber \\
\A \A
	\hspace{1cm}\left.
	-4i(\bar{\zeta}\sigma^{cd}\sigma^{ab}\sigma^{ef}\gamma_c\tilde{\lambda}{}_{d})\omega_{ef\mu}
	+2(\bar{\zeta}\sigma^{ef}\sigma^{ab}\gamma^d\tilde{\lambda}{}_{d})\omega_{ef\mu}
	\right]
	\nonumber \\
\A \A
+\frac{1}{4\kappa}
	\left[
	-4(\bar{\zeta}\sigma^{[b|c}\tilde{\Lambda}^{d|}{}_{d})\omega^{a]}{}_{c\mu}
	+i(\bar{\zeta}\sigma^{ab}\sigma^{cd}\tilde{\Lambda}^{e}{}_{e})\omega_{cd\mu}
	-(\bar{\zeta}\sigma^{cd}\sigma^{ab}\sigma^{ef}{\tilde{\Lambda}}_{cd})\omega_{ef\mu}
	\right.
	\nonumber \\
\A \A
	\hspace{1cm}\left.
	-(\bar{\zeta}\sigma^{c\rho}\sigma^{ab}\sigma^{de}\tilde{\Lambda}_{(c\mu )})\omega_{de\rho}
	-2(\bar{\zeta}\sigma^{ef}\sigma^{cd}\sigma^{ab}{\tilde{\Lambda}}_{cd})\omega_{ef\mu}
	\right], 
\label{lsusy-compo-5}
\ea
where $\epsilon^{ab}$ is the Lorents parameter and we put 
$\epsilon^{ab}=\xi^{\rho}\omega^{ab}{}_{\rho}$. 
$\delta B_{\mu}$ and $\delta B{^a}_{\mu}$ are similar to $\delta A_{\mu}$ and $\delta A{^a}_{\mu}$ 
respectively and omitted for simplicity. 
In the right-hand side of (\ref{lsusy-compo-3}) and $\delta B{^a}_{\mu}$, 
the last terms contain $\Lambda^{\prime}{}^a{}_{\mu}$ which is defined by 
$\Lambda^{\prime}{}^a{}_{\mu}=-\epsilon^{abcd}\gamma_5\psi\omega_{bcd}$ . 
Note that $\Lambda^{\prime}{}^a{}_{\mu}$ is not the functional of 
the supermultiplet (\ref{80fermions1}), 
so we may have to treat $\Lambda^{\prime}{}^a{}_{\mu}$ as new auxiliary field. 
However, if we put 
$\epsilon^{ab}=\epsilon^{ab}(\tilde{\lambda}{}_{\mu}, \tilde{\Lambda}{}^a{}_{\mu})$, e.g. 
$\epsilon^{ab}=\bar\zeta\gamma^{[a}\tilde{\lambda}{}^{b]}$, 
$\Lambda^{\prime}{}^a{}_{\mu}$ does not appear in 
the right-hand side of (\ref{lsusy-compo-3}) and $\delta B^{a}{_\mu}$. 
As a result, the LSUSY transformation 
on the supermultiplet (\ref{80bosons1}) and (\ref{80fermions1}) 
are written by using the supermultiplet itself 
at least at the leading order of superon $\psi$. 
The higher order terms remain to be studied.    
However we believe that we can obtain the complete linearized off-shell supermultiplets of 
the SP algebra by repeating the similar procedures (on the auxiliary fields) order by order 
which terminates with $\psi^{4}$.  It may be favorable that  10 bosonic auxiliary fields, for example 
${a_{\mu}(e,\psi), b_{\mu}(e,\psi), M(e,\psi), N(e,\psi)}$ are arbitrary  up now and available 
for the closure of the off-shell SP algebra in higher order terms.     \par
Here we just mention the systematic linearization by using the superfield formalism 
applied to study the coupled system of VA action with SUGRA\cite{lr}. 
We can define on such a coupled system 
a local spinor gauge symmetry which induces a super-Higgs mechanism\cite{dz1} converting VA field to 
the longitudinal component of massive spin 3/2 field. The consequent Lagrangian  
may be an analogue  that we have  anticipated in the composite picture 
but with the elementary spin 3/2 field. 
Developing the superfield formalism on SGM spacetime may be crucial  for carrying out 
the linearization along the SGM composite scenario, especially for $N>1$.    \par
Finally, we discuss the commutators for more general cases.   \\
Here we consider a functional of $(e{^a}_\mu, \psi)$ and their derivatives as 
\ba
f_A(\psi, \bar\psi, e{^a}_\rho; 
\psi_{,\rho}, \bar\psi_{,\rho}, e{^a}_{\rho,\sigma}),  \ (A = \mu, \mu \nu, ... etc.) 
\ea
with $\psi_{,\rho} = \partial_\rho \psi, etc.$, 
and we suppose that $f_A$ is the functional of $O(\psi^2)$ for simplicity. 
Then we have the variation of $f_A$, 
\ba
\delta f_A = {{\partial f_A} \over {\partial \psi}} \delta \psi 
+ \delta \bar\psi {{\partial f_A} \over {\partial \bar\psi}} 
+ {{\partial f_A} \over {\partial e{^a}_\rho}} \delta e{^a}_\rho 
+ {{\partial f_A} \over {\partial \psi_{,\rho}}} (\delta \psi)_{,\rho} 
+ (\delta \bar\psi)_{,\rho} {{\partial f_A} \over {\partial \bar\psi_{,\rho}}} 
+ {{\partial f_A} \over {\partial e{^a}_{\rho,\sigma}}} 
(\delta e{^a}_\rho)_{,\sigma}.  
\ea
and the commutator for $f_A$ becomes 
\ba
[\delta_1, \delta_2] f_A 
= \A \A {{\partial f_A} \over {\partial \psi}} [\delta_1, \delta_2] \psi 
+ [\delta_1, \delta_2] \bar\psi {{\partial f_A} \over {\partial \bar\psi}} 
+ {{\partial f_A} \over {\partial e{^a}_\rho}} [\delta_1, \delta_2] e{^a}_\rho 
\nonu
\A \A + {{\partial f_A} \over {\partial \psi_{,\rho}}} 
([\delta_1, \delta_2] \psi)_{,\rho} 
+ ([\delta_1, \delta_2] \bar\psi)_{,\rho} 
{{\partial f_A} \over {\partial \bar\psi_{,\rho}}} 
+ {{\partial f_A} \over {\partial e{^a}_{\rho,\sigma}}} 
([\delta_1, \delta_2] e{^a}_\rho)_{,\sigma} 
\label{com-f}
\ea
If we substitute the commutators for $(e{^a}_\mu, \psi)$ of Eq.(8) 
into Eq.(\ref{com-f}), we obtain 
\begin{equation}
[\delta_1, \delta_2] f_A = \Xi^\lambda \partial_\lambda f_A + G_A, 
\label{com2-f}
\end{equation}
where $G_A$ is defined by 
\ba
G_A = \A \A \partial_\rho \Xi^\lambda 
\left( {{\partial f_A} \over {\partial e{^a}_\rho}} e{^a}_\lambda 
+ {{\partial f_A} \over {\partial \psi_{,\rho}}} \partial_\lambda \psi 
+ \partial_\lambda \bar\psi 
{{\partial f_A} \over {\partial \bar\psi_{,\rho}}} 
+ {{\partial f_A} \over {\partial e{^a}_{\sigma,\rho}}} 
\partial_\lambda e{^a}_\sigma 
+ {{\partial f_A} \over {\partial e{^a}_{\rho,\sigma}}} 
\partial_\sigma e{^a}_\lambda \right) 
\nonu
\A \A + \partial_\rho \partial_\sigma \Xi^\lambda 
{{\partial f_A} \over {\partial e{^a}_{\rho,\sigma}}} e{^a}_\lambda. 
\label{GA}
\ea
The first term in r.h.s. of Eq.(\ref{com2-f}) means the translation of $f_A$. 
Therefore Eq.(\ref{com2-f}) shows that the closure of the commutator algebra 
on GL(4,R) for the various functionals $f_A$ in the previous argument 
depends on $G_A$ of Eq.(\ref{GA}), 
and these argument reproduces all the previous commutators respectively.  \par
The linearization of SGM action (\ref{SGM1}) with the extra dimensions, 
which gives another unification framework describing the observed particles as elementary fields, is open. 
And the linearization of SGM action for spin 3/2 NG fermion field\cite{st3} discussed in the next section 
(with extra dimensions) 
to be discussed in the next section may be in the same scope.  \par
Now we summarize the results as follows: 
(i) Referring to SUGRA transformations we have obtained explicitly the SUSY invariant 
relations up to $O(\psi)^{2}$ and the corresponding new LSUSY transformations 
among 80+80 off-shell supermultiplet of LSUSY. 
(ii) The new LSUSY transformations on 80+80 linearized supermultiplet are different apparently 
from SUGRA transformations but close on super-Poincar\'e. 
(iii)It is interesting that the simple relation $\lambda_{\mu}=e^{a}{_\mu}\gamma_{a}\psi + \cdots $, 
which is sugested by the flat spacetime linearization, seems disfavour with the SGM linearization 
in our present method, so far. 
From the physical viewpoint  what LSUSY SP may be to SGM in quantum field theory, 
what O(4) symmetry is to the relativistic hydrogen model in quantum mechanics. 
The complete linearization to all orders up to $O(\psi)^{4}$, 
which can be anticipated by the systematics emerging in the present study, needs specifications of 
the auxiliary fields and remains to be studied. The details will appear separately\cite{sts4}. 
\section{SGM with spin 3/2 Superon}
In this section  we extend the SGM\cite{ks3} to a higher spin  NG fermion. 
Following the arguments of VA, the action of NG fermion $\psi^{\mu}_{\alpha}(x)$ with spin 3/2 
is already written down by Baaklini as a nonlinear realization of a new superalgebra containing 
a vector-spinor generator $Q^{\mu}_{\alpha}$\cite{b}.
We study in detail the gravitational interaction of Baaklini model\cite{b}. 
We will see that the similar  arguments to  SGM can be performed and produce a  gauge invariant action, 
which is the straightforward generalization of SGM action. The phenomenological implications of spin 3/2 
fundamental constituents are discussed briefly. 

In ref.\cite{b}, a new SUSY algebra containing a spinor-vector generator $Q^{\mu}_{\alpha}$ is introduced as follows:
\begin{equation}
\{Q^{\mu}_{\alpha},Q^{\nu}_{\beta}\}=\varepsilon^{\mu\nu\lambda\rho}P_{\lambda}(\gamma_{\rho}\gamma_{5}C)_{\alpha\beta}, 
\label{qq}
\end{equation} 
\begin{equation}
[Q^{\mu}_{\alpha},P^{\nu}]=0,
\label{qp}
\end{equation} 
\begin{equation}
[Q^{\mu}_{\alpha},J^{\lambda\rho}]={1\over2}(\sigma^{\lambda\rho}Q^{\mu})_{\alpha}+
i\eta^{\lambda\mu}Q^{\rho}_{\alpha}-i\eta^{\rho\mu}Q^{\lambda}_{\alpha},
\label{qj}
\end{equation} 
where $Q^{\mu}_{\alpha}$ are vector-spinor generators satisfying Majorana condition 
$Q^{\mu}_{\alpha}=C_{\alpha\beta}{\overline Q}^{\mu}_{\alpha}$, $C$ is a charge conjugation matrix 
and ${1 \over 2}\{\gamma^{\mu},\gamma^{\nu}\}=\eta^{\mu\nu}=(+,-,-,-)$.
By extending  the arguments of VA model\cite{va}of NLSUSY, they obtain the following action 
as the nonlinear representation of the new SUSY algabra. 
\begin{equation}
S={1 \over \kappa}\int \omega_{0} \wedge \omega_{1} \wedge \omega_{2} \wedge \omega_{3}
={1 \over \kappa}\int \det{w_{ab}}d^{4}x,
\label{va3/2}
\end{equation} 
\begin{equation}
w_{ab}={\delta}_{ab}+ t_{ab},  \quad
t_{ab}=i\kappa \varepsilon_{acde}\bar{\psi}^{c}\gamma^{d}\gamma_{5}\partial_{b}{\psi}^{e},
\label{bak-wt}
\end{equation} 
where $\kappa$ is  up to now  arbitrary constant with the dimension of the fourth power of length(i.e., 
a fundamental volume of  spacetime) 
and  $\omega_{a}$ is the following differential forms
\begin{equation}
\omega_{a} =dx_{a} + i \kappa \varepsilon_{abcd}\bar{\psi}^{b}\gamma^{c}\gamma_{5}d{\psi}^{d},
\label{bak-omega}
\end{equation} 
which is invariant under the following (super)translations 
\begin{equation}
\psi^{a}_{\alpha} \longrightarrow   \psi^{a}_{\alpha} + \zeta^{a}_{\alpha}, 
\label{bak-3/2}
\end{equation} 
\begin{equation}
x_{a}  \longrightarrow   x_{a} + i\kappa \varepsilon_{abcd}\bar{\psi}^{b}\gamma^{c}\gamma_{5}{\zeta}^{d}, 
\label{bak-x}
\end{equation} 
where $\zeta^{a}_{\alpha}$ is a constant Majorana tensor-spinor parameter.

Now we consider the  gravitational interaction of Baaklini model(\ref{qq}). 
We show that the arguments performed in  SGM of spin 1/2  NG field $\psi_{\alpha}(x)$ 
can be extended straightforwardly to spin 3/2 Majorana NG field $\psi^{a}_{\alpha}$. 
In the present case, as seen in (\ref{bak-omega}), (\ref{bak-3/2}) and  (\ref{bak-x}) 
NLSUSY SL(2C) degrees of freedom (i.e. the coset space coordinates  $\psi^{a}_{\alpha}$ 
representing NG fermions) in addition to 
Lorentz SO(3,1) coordinates are embedded at every curved spacetime point with GL(4R) invariance.    \\
Following the arguments of SGM\cite{ks3}, it is natural to introduce formally a new vierbein field ${w^{a}}_{\mu}(x)$ through the NLSUSY invariant differential forms 
$\omega_{a}$ in (\ref{bak-omega})  as follows: 
\begin{equation}
\omega^{a} = {w^{a}}_{\mu}dx^{\mu}, 
\label{3/2SGM-omega}
\end{equation} 
\begin{equation}
{w^{a}}_{\mu}(x) = {e^{a}}_{\mu}(x) +  {t^{a}}_{\mu}(x),    \quad  
{t^{a}}_{\mu}(x)=i\kappa \varepsilon^{abcd}\bar{\psi}_{b}\gamma_{c}\gamma_{5}\partial_{\mu}{\psi}_{d}, 
\label{3/2SGM-w-t}
\end{equation} 
where ${e^{a}}_{\mu}(x)$ is the vierbein of Einstein Genaral Relativity Theory(EGRT) and 
Latin $(a,b,..)$ and Greek $(\mu,\nu,..)$ are the indices for local Lorentz and general coordinates, respectively.
By noting $(\psi^{\mu}_{\alpha}(x))^{2}=0$, we can easily obtain the inverse of  the new vierbein, ${w_{a}}^{\mu}(x)$, 
in the power series of ${t^{a}}_{\mu}$ which terminates with $({t^{a}}_{\mu})^{4}$: 
\begin{equation}
{w^{\mu}{_a}} = e^{\mu}{_a} - t^{\mu}{_a} +{t^{\rho}}_{a}{t^{\mu}}_{\rho}- \dots . 
\label{3/2SGM-w-inverse}
\end{equation}  
Note that the first and the second indices of ${t^{a}}_{\mu}$ ($t^{\mu}{_a}$) represent 
those of $\gamma$-matrix and the derivative, respectively.
Similarly we introduce formally a new metric tensor  $s^{\mu\nu}(x)$ in the abovementioned curved spacetime 
as follows:
\begin{equation}
s^{\mu\nu}(x) \equiv {w_{a}}^{\mu}(x) w^{{a}{\nu}}(x). 
\label{3/2SGM-metric}
\end{equation}
It is easy to show 
${w_{a}}^{\mu} w_{{b}{\mu}} = \eta_{ab}$,  $s_{\mu \nu}{w_{a}}^{\mu} {w_{b}}^{\mu}= \eta_{ab}$, ..etc. 
In order to obtain simply the action in the abovementioned curved spacetime, which is invariant  at least 
under GL(4R), NLSUSY and local Lorentz transformations, we follow formally EGRT as performed in SGM. 
That is, we require that the new unified vierbein ${w^{a}}_{\mu}(x)$ and the  metric  
$s^{\mu\nu}(x)$ should have formally a general coordinate transformations  under the supertranslations: 
\begin{equation}
\delta x_{\mu}= - \xi_{\mu}, \quad 
\delta\psi^{a}=\zeta^{a},
\label{3/2SGM-trans}
\end{equation}
where $\xi^{\mu}=
i\kappa \varepsilon^{\mu\nu\rho\sigma}\bar{\psi}_{\nu}\gamma_{\rho}\gamma_{5}{\zeta}_{\sigma}$.    \\ 
Remarkably we find that the following global new NLSUSY transformations 
\begin{equation}
\delta \psi^{a}(x) = \zeta^{a} - i\kappa (\varepsilon^{\mu\nu\rho\sigma}\bar{\psi}_{\nu}\gamma_{\rho}
\gamma_{5}{\zeta}_{\sigma})\partial_{\mu}\psi^{a}
\label{3/2SGM-spintrans}
\end{equation} 
\begin{equation}
\delta {e^{a}}_{\mu}(x) = i\kappa (\varepsilon^{\rho\nu\sigma\lambda}\bar{\psi}_{\nu}\gamma_{\sigma}
\gamma_{5}{\zeta}_{\lambda})\partial_{[\mu}{e^{a}}_{\rho]}
\label{3/2SGM-vierbeintrans}
\end{equation} 
induce the desirable transformations on ${w^{a}}_{\mu}(x)$ and $s^{\mu\nu}(x)$ as follows: 
\begin{equation}
\delta_{\zeta_{1}} {w^{a}}_{\mu} = \xi^{\nu}_{1} \partial_{\nu}{w^{a}}_{\mu} + \partial_{\mu} \xi^{\nu}_{1} {w^{a}}_{\nu}, 
\end{equation} 
\begin{equation}
\delta_{\zeta_{1}} s_{\mu\nu} = \xi^{\kappa}_{1} \partial_{\kappa}s_{\mu\nu} +  
\partial_{\mu} \xi^{\kappa}_{1} s_{\kappa\nu} 
+ \partial_{\nu} \xi^{\kappa}_{1} s_{\mu\kappa}, 
\end{equation} 
where  $\xi^{\rho}_{\zeta_{1}}=
i\kappa \varepsilon^{\mu\nu\rho\sigma}\bar{\psi}_{\nu}\gamma_{\rho}\gamma_{5}{\zeta}_{1\sigma}$. 
These show that ${w^{a}}_{\mu}(x)$ and $s^{\mu\nu}(x)$  have general coordinate transformations under 
the new NLSUSY transformations (\ref{3/2SGM-spintrans}) and (\ref{3/2SGM-vierbeintrans}).       \\
Therefore replacing  ${e^{a}}_{\mu}(x)$ in EH Lagrangian of general relativity 
by  the new vierbein ${w^{a}}_{\mu}(x)$  
we obtain the following Lagrangian which is invariant under (\ref{3/2SGM-spintrans}) and 
(\ref{3/2SGM-vierbeintrans}):
\begin{equation}
L=-{c^{3} \over 16{\pi}G}\vert w \vert(\Omega + \Lambda ),
\label{3/2SGM}
\end{equation}
\begin{equation}
\vert w \vert=det{w^{a}}_{\mu}=det({e^{a}}_{\mu}+{t^{a}}_{\mu}),  
\end{equation} 
where the overall factor is now fixed uniquely to ${-c^{3} \over 16{\pi}G}$, 
${e_{a}}^{\mu}(x)$ is the vierbein of EGRT and 
$\Lambda$ is a probable cosmological constant. 
$\Omega$ is a (mimic) new  unified scalar curvature analogous to the Ricci scalar curvature $R$ of EGRT. 
The explicit 
expression of $\Omega$ is obtained  by just replacing ${e_{a}}^{\mu}(x)$ in Ricci scalar $R$ of EGRT by 
${w_{a}}^{\mu}(x)={e^{a}}_{\mu}+{t^{a}}_{\mu}$, which gives the gravitational interaction of  $\psi^{a}_{\alpha}(x)$. 
The lowest order term of $\kappa$ in the action (\ref{3/2SGM}) gives the EH action 
of general relativity. 
And in flat spacetime, i.e.  ${e_{a}}^{\mu}(x) \rightarrow {\delta_{a}}^{\mu}$, 
the action (\ref{3/2SGM}) reduces to  VA model with  ${\kappa}^{-1} = {c^{3} \over 16{\pi}G}{\Lambda}$. 
Therefore our model predicts  a non-zero (small) cosmological constant.             \\
As for the Lorentz invariance we again require that the new vierbein 
$w{^{a}}_{\mu}(x)$ should have formally 
a local Lorentz transformation as for SGM with spin 1/2 NG fermion. 
Then we find that the following (generalized) 
local Lorentz transformations 
\ba
\A \A \delta_L \psi^a(x) = \epsilon{^a}_b \psi^b 
      - {i \over 2} \epsilon_{bc} \sigma^{bc} \psi^a, 
\label{Lr3/2-p} \\
\A \A \delta_L e{^a}_{\mu}(x) = \epsilon{^a}_b e{^b}_{\mu} 
      - i\kappa \varepsilon^{abcd} 
      \{ \bar{\psi}_b \gamma_c \gamma_5 \psi_e 
      (\partial_{\mu} \epsilon{_d}^e) 
      - {i \over 4} \varepsilon{_c}^{efg} 
      \bar{\psi}_b \gamma_g \psi_d 
      (\partial_{\mu} \epsilon_{ef}) \} 
\label{Lr3/2-e}
\ea
induce the desirable transformation. 
The equation (\ref{Lr3/2-e}) also reduces 
to the familiar form of the Lorentz transformations 
if the global transformations are considered ( for $g^{\mu\nu}$ ). 

Therefore, as in spin 1/2 SGM case,  replacing $e{^a}_{\mu}(x)$ 
in EH Lagrangian of GR 
by the new vierbein $w{^a}_{\mu}(x)$ defined by (\ref{3/2SGM-w-t}), 
we obtain the Lagrangian (\ref{3/2SGM}) of the same form as (\ref{SGM}), 
which is invariant under (\ref{3/2SGM-spintrans}), (\ref{3/2SGM-vierbeintrans}), 
(\ref{Lr3/2-p}) and (\ref{Lr3/2-e}). 

The commutators of two new supersymmetry transformations 
(\ref{3/2SGM-spintrans}) and (\ref{3/2SGM-vierbeintrans})
on $\psi^{a}(x)$ and  $e{_a}^{\mu}(x)$ 
are now calculated as\cite{st1} 
\ba
\A \A [\delta_{\zeta_1}, \delta_{\zeta_2}] \psi^{a} 
      = \{ 2i\kappa (\varepsilon^{\mu bcd} \bar{\zeta}_{2b} 
      \gamma_c \gamma_5 \zeta_{1d}) 
      - \xi_1^{\rho} \xi_2^{\sigma} e{_a}^{\mu} 
      (\partial_{[\rho} e{^a}_{\sigma]}) \} 
      \partial_{\mu} \psi^a, \\
\A \A [\delta_{\zeta_1}, \delta_{\zeta_2}] e{^a}_{\mu} 
      = \{ 2i\kappa (\varepsilon^{\rho bcd} \bar{\zeta}_{2b} 
      \gamma_c \gamma_5 \zeta_{1d}) 
      - \xi_1^{\sigma} \xi_2^{\lambda} e{_c}^{\rho} 
      (\partial_{[\sigma} e{^c}_{\lambda]}) \} 
      \partial_{[\rho} e{^a}_{\mu]} \nonu
\A \A \hspace{2.5cm} 
      - \partial_{\mu} (\xi_1^{\rho} \xi_2^{\sigma} 
      \partial_{[\rho} e{^a}_{\sigma]}). 
\ea
These can be rewritten as GL(4R);
\ba
\A \A [\delta_{\zeta_1}, \delta_{\zeta_2}] \psi^{a} 
      = \Xi^{\mu} \partial_{\mu} \psi^{a}, 
\label{com3/2-p} \\
\A \A [\delta_{\zeta_1}, \delta_{\zeta_2}] e{^a}_{\mu} 
      = \Xi^{\rho} \partial_{\rho} e{^a}_{\mu} 
      + e{^a}_{\rho} \partial_{\mu} \Xi^{\rho}, 
\label{com3/2-e}
\ea
where $\Xi^{\mu}$ is now a generalized gauge parameter defined by 
\begin{equation}
\Xi^{\mu} = 2i\kappa (\varepsilon^{\mu bcd} \bar{\zeta}_{2b} 
      \gamma_c \gamma_5 \zeta_{1d}) 
      - \xi_1^{\rho} \xi_2^{\sigma} e{_a}^{\mu} 
      (\partial_{[\rho} e{^a}_{\sigma]}). 
\end{equation}
Also, the commutator of 
the local Lorentz transformation on $e{_a}^{\mu}(x)$ 
of Eq.(\ref{Lr3/2-e}) is calculated as 
\begin{equation}
[\delta_{L_{1}}, \delta_{L_{2}}] e{^a}_{\mu} 
= \beta{^a}_b e{^b}_{\mu} 
- i\kappa \varepsilon^{abcd} 
\{ \bar{\psi}_b \gamma_c \gamma_5 \psi_e 
(\partial_{\mu} \beta{_d}^e) 
- {i \over 4} \varepsilon{_c}^{efg} 
\bar{\psi}_b \gamma_g \psi_d 
(\partial_{\mu} \beta_{ef}) \} 
\label{comLr3/2}
\end{equation}
where $\beta_{ab}$ is the same as SGM with spin 1/2. 
The equations (\ref{Lr3/2-e}) and (\ref{comLr3/2}) 
explicitly reveal a generalized local Lorentz transformation 
with the parameters $\epsilon_{ab}$ and $\beta_{ab}$, 
which forms a closed algebra.     \\
Therefore our action (\ref{3/2SGM}) is invariant at least under   
\begin{equation}
[{\rm global\ new \ NLSUSY}] \otimes [{\rm local\ GL(4,R)}] 
\otimes [{\rm local\ Lorentz}] \otimes [{\rm global\ SO(N)}],
\end{equation}
when it is extended to global SO(N). 
It is interesting that the spin 3/2 massless field can couple consistently with graviton 
besides SUGRA.
SGM formalism \cite{ks3} can be  generalized 
to the spacetime with extra dimensions for the inclusion of the non-abelian internal symmetries. 
It may give a potential new framework for the simple unification of spacetime and matter.   \\
Finally we just mention the phenomenological implications of our model. 
As read off from the above discussions it is easy to introduce (global) SO(N) internal symmeytry in our model 
by replacing $\psi^{a}_{\alpha}(x) \rightarrow  {\psi^{ia}}_{\alpha}(x),(i=1,2, \dots, N)$,  
which may enable us to consider  SGM[5] with spin 3/2 superon. However the fundamental internal symmetry 
for superons may be rather different from SGM, for the generator of a new algebra shifts spin by 3/2 
and one-superon states correspond to spin 1/2 states but the adjoint representation is a vector as well. 
Also it is worthwhile to consider SGM with  extra dimensions.   
We think that the above result is useful when we consider the gravitational interaction of 
the massless field with higher half-integer spin $(> 5/2)$, though the algebra itself contains 
the negative norm states\cite{s-ta}. 
\section{Cosmology of New EH-type Action}
There remain many unsolved interesting problems, even qualitatively, in the physics of the universe, e.g. 
the birth of the universe which is expanding, the origins of the inflation and the big bang, 
the tiny value of the cosmological constant, the critical value$(\sim 1)$ of the energy density,  
the dark energy, the baryon number genesis, ... etc. 
These problems should be understood in terms of  the knowledges of the unified local field theory of 
particle physics.    
We discuss briefly and qualitatively the potential of SGM for these unsolved problems.            \\
We regard that the ultimate entity of nature is  high symmetric SGM spacetime inspired by NLSUSY, 
where  the coset space coordinates $\psi$ of ${superGL(4,R) \over GL(4,R)}$ 
turning to the NG fermion d.o.f. in addition to the ordinaly Minkowski coordinate $x^{a}$, 
i.e.  $ local \ SL(2C) \times local \ SO(3,1)$ d.o.f., are attached at every  spacetime point.  
The geometry of new spacetime is described by SGM action (\ref{SGM}) of {\it vacuum \ EH-type} 
and gives the unified description of nature. 
The fundamental action  (\ref{SGM}) on new (SGM) spacetime is  unstable against 
the new {\it global} NLSUSY transformation and induces 
{\it the self-contained spontaneous (symmetry) breakdown} into  ordinary observed  Riemann spacetime 
and the massless superon-quintet matter which  expands rapidly, 
for the cuvature-energy of SGM  spacetime is converted into those of Riemann spacetime 
and energy-momentum of the superon(matter). 
This may be regarded as the gapless phase transition of spacetime from SGM to Riemann.
Also this may be the birth of the present expanding universe, i.e. the big bang and 
the consequent rapid expansion (inflation) of spacetime in the quark-lepton SM era.  
And we think that the birth of the universe  by the {\it spontaneous \ breakdown} 
of self-contained SGM spacetime  of {\it vacuum} action of EH-type (\ref{SGM}) 
may explain qualitatively the observed critical value$(\sim 1)$  of the energy density 
of the universe.   \\ 
Note that SGM action poseses  two inequivalent flat spaces, 
one is SGM-flat($w{^a}_{\mu}(x) \rightarrow {\delta}{^a}_{\mu}$) 
which allows nontrivial configurations of $e^{a}{_\mu}$ and $t^{a}{_\mu}$  and 
the other is Riemann flat($e{^a}_{\mu}(x) \rightarrow {\delta}{^a}_{\mu}$), 
which are crucial for the spontaneous breakdown from SGM to Riemann sapcetime. 
As proved for EH action of  GR~\cite{wttn}, the energy of SGM action of 
EH-type is expected to be  positive (for positive $\Lambda$).  \par
Remarkably the observed Riemann spacetime of
EGRT and matter(superons) appear  simultaneously from (the vacuum
) SGM action by the spontaneous decay of SGM spacetime, 
i.e. by the gapless phase transition of spacetime.  
The  catastrophe problem of the gravitational collapse  should be reconsidered in SGM due to the 
massless NG mode at the Planck scale, i.e. the phase transition to and subsequently from the unstable 
SGM spacetime.    \\
It is interesting if SGM could give new insights into these unsolved problems.   \par
\section{Discussions}
A new EH-type action( called tentatively SGM from the composite viewpoints ) 
in NLSUSY inspired (SGM) spacetime is obtained by the geometrical arguments similar to 
Einstein general relativity theory in Riemann spacetime. 
Despite the simple expressions  of the  unified vierbein defined on N=1 SGM spacetime 
${w^{a}}_{\mu} ={e^{a}}_{\mu}+ {t^{a}}_{\mu}$, 
${w^{\mu}{_a}}= e^{\mu}{_a}- t{^{\mu}}_a + {t^{\rho}}_a {t^{\mu}}_{\rho} 
- t{^{\rho}}_a t{^{\sigma}}_{\rho} t{^{\mu}}_{\sigma} 
+ t{^{\rho}}_a t{^{\sigma}}_{\rho} t{^{\kappa}}_{\sigma}t{^{\mu}}_{\kappa}$, 
(Note that the second index of $t$ represents the derivative.)
and the consequent metric $s_{\mu\nu}, \cdots,$ etc, \ SGM is a nontrivial generalization 
of EH action.     
In fact, as for the bosonic gauge transformation we can show explicitly that 
by the redefinitions(variations under GL(4R) with the field dependent parameters) 
only on the vierbein, e.g. 
${e^{a}}_{\mu} \rightarrow {e^{a}}_{\mu}-{t^{a}}_{\mu}$ 
and the consequent variations on  $e^{\mu}{_a}$ it is impossible to gauge away $\psi$ 
in compatible with new NLSUSY, 
for the new NLSUSY induces the square-root of GL(4R) on $w$ ( and $s$)  and 
defined on the {\it multiplet} $( \ {e^{a}}_{\mu},\  \psi \ )$.  \\
Next we discuss on  the confusive local spinor transformation which leaves SGM action invariant. 
SGM action (\ref{SGM}) is invariant under  
the following local spinor translation with  a local parameter $\epsilon(x)$, 
$\delta\psi=\epsilon$ and 
$\delta {e^{a}}_{\mu}=
-i \kappa^{2}( \bar\epsilon\gamma^{a}\partial_{\mu}\psi+\bar\psi\gamma^{a}\partial_{\mu}\epsilon)$ 
which give  $\delta {w^{a}}_{\mu} = \delta {w^{\mu}{_a}} = 0$.  
It should be noticed that the local fermionic d.o.f. $\psi$ would not be transformed (gauged) away. 
At a glance, the choice $\delta\psi=\epsilon= -\psi$ seemingly gauges away $\psi$. 
However, such an effect is canceled precisely by the simultaneous gauge transformation 
$\delta {e^{a}}_{\mu}$ with $\epsilon= -\psi$, 
i.e. 
$w(e,\psi)=w(e + \delta e,\ \psi +  \delta \psi) = w( e  + t, \ 0  )$ as indicated by $\delta w=0$, 
which reproduces precisely SGM action describing NLSUSY invariant gravitational interaction of superon.  
The commutators of these local spinor transformations on ${e^{a}}_{\mu}$ and $\psi$  vanish 
identically. 
Therefore the local spinor translation  mentioned above is a fake (gauge) transformation 
in a sence that, in contrast with the local Lorentz transformation invariance 
in EGRT, it can not eliminate gauge d.o.f. with spin 1/2, 
for the unified vierbein $w= e + t$ is the only gauge field on SGM spacetime that 
contains only integer spin. 
(Note that the puzzling (spacetime origin) local spinor symmetry  plays an essential role 
in the linearization, i.e. in the superHiggs  phenomena\cite{dz} as demonstrated 
in the  coupled system of VA action of NLSUSY and SUGRA of LSUSY  equipted 
with a mass term and a cosmological constant.) 
This confusive situation comes from the funny geometrical formulation of SGM on unfamiliar SGM spacetime 
where, besides the Minkowski coordinates $x^{a}$,  $\psi$ is a Grassmann {\it  coordinate} 
(i.e., the another fundamental d.o.f.) defining the tangential spacetime with $SO(3,1) \times SL(2C)$ d.o.f. 
inspired by NLSUSY. (Note that  $SO(3,1)$ is the twice covering group of $SL(2C)$.) 
And $\delta \psi = \epsilon(x)$ is just a coordinate translation(redefinition) 
on SGM flat spacetime. 
These situations can be understood easily by observing that the unified vierbein gauge field 
${w^{a}}_{\mu}(x) ={e^{a}}_{\mu}(x)+ {t^{a}}_{\mu}(x)$  is defined   by 
$\omega^{a}=dx^{a} + {i \kappa^{4}  \over 2}(\bar{\psi}^{j}\gamma^{a}
d{\psi}^{j}
- d{\bar{\psi}^{j}}\gamma^{a}{\psi}^{j})
\sim {w^{a}}_{\mu}dx^{\mu}$, 
where $\omega^{a}$ is the NLSUSY invariant differential form of VA and that $x^{a}$ and $\psi$ are 
coordinates of flat spacetime inspired by NLSUSY and SGM are encoded as a spacetime symmetry.  
From these geometrical viewpoints (in SGM spacetime)  we can  understand that 
$\psi$ is a coordinate and would be neither transformed away nor gauge-fixed away 
and the structure of SGM (flat) spacetime is preserved.  
Note that putting $\psi=0$ by formal arguments concerning the local spinor symmetry 
makes SGM based upon NLSUSY vacuum(VA flat spacetime action) reduce to EH action 
based upon different vacuum(Minkowski flat spacetime), which is another theory based upon another 
vacuum. 
SGM (\ref{SGM}) is  a nontrivial generalization of EH action and Born-Infeld action\cite{bi}.  \\
SGM posesses some hidden global symmetries originating from the fact that the graviton 
and (the energy-momentum  of) the superon 
contribute equally to the unified vierbein $w$\cite{st1}.    \par
In this talk we have presented an attempt to describe the unity of nature as a geometry 
of new spacetime manifold with high symmetry and rich structures, 
which is called tentatively SGM spacetime from the viewpoints of the compositeness of matter.  
New (SGM) spacetime is the ultimate physical entity described by EH-type vacuum action (\ref{SGM}) 
and induces  spontaneously the  phase transition  to observed Riemann spacetime and matter. 
We have depicted  the potential of new EH-type (SGM) action. 
The study of the vacuum structure of SGM action in the broken phase
(i.e. SGM action in Riemann spacetime) is important and challenging.    \par
SGM  with the extra dimensions to be compactified is open, which may allow the unification by means of the 
elementary fields.  In this case the mechanism of the conversion of the spacetime d.o.f. 
into the dynamical d.o.f. is duplicate, i.e. by the compactification of Kaluza-Klein type and 
by the new mechanism adopted in SGM.  \\
SGM with spin 3/2 NG fermion may be in the same scope but remains to be studied. 
\newpage

One of the authors(K.S.) would like to thank Professor M. Rashdan 
for inviting him to 
the first big International conference on this field in Egypt held at the 
Tibarose({\it  Rose of Tehbe} ) and for his warm hospitality  and 
Professor W. Greiner  for encouraging discussions and  for the invitation 
to the publication of the talk in the journal.    
Also he would like to thank Professor J. Wess for his encouragement, enlightening discussions 
and the warm hospitality  through the researches.
We are  also grateful to Dr. T. Matsuda for discussions on the cosmology. 

\newpage

%
\newcommand{\NP}[1]{{\it Nucl.\ Phys.\ }{\bf #1}}
\newcommand{\PL}[1]{{\it Phys.\ Lett.\ }{\bf #1}}
\newcommand{\CMP}[1]{{\it Commun.\ Math.\ Phys.\ }{\bf #1}}
\newcommand{\MPL}[1]{{\it Mod.\ Phys.\ Lett.\ }{\bf #1}}
\newcommand{\IJMP}[1]{{\it Int.\ J. Mod.\ Phys.\ }{\bf #1}}
\newcommand{\PR}[1]{{\it Phys.\ Rev.\ }{\bf #1}}
\newcommand{\PRL}[1]{{\it Phys.\ Rev.\ Lett.\ }{\bf #1}}
\newcommand{\PTP}[1]{{\it Prog.\ Theor.\ Phys.\ }{\bf #1}}
\newcommand{\PTPS}[1]{{\it Prog.\ Theor.\ Phys.\ Suppl.\ }{\bf #1}}
\newcommand{\AP}[1]{{\it Ann.\ Phys.\ }{\bf #1}}


\begin{thebibliography}{100}
%
\bibitem{wz} J. Wess and B. Zumino, {\it Phys. Lett.} {\bf B49}, 52 (1974).  
%
\bibitem{va}  D.V. Volkov and V.P. Akulov, {\it Phys. Lett.} {\bf B46}, 109(1973). 
%
\bibitem{gl}  Y.A. Golfand and E.S. Likhtman, {\it JET. Lett.} {\bf 13}, 323 (1971).          
%
\bibitem{fnf}  D.Z. Freedman, P. van Nieuwenhuizen and S. Ferrara, \PR{D13}, 3214(1976). 
%
\bibitem{dz}  S. Deser and B. Zumino, \PL{B62} (1976) 335.
%
\bibitem{dn}  B. de Wit and H. Nicolai,  {\it Phys. Lett.} {\bf B108}, 285(1982).
%
\bibitem{g}   M. Gell-Mann, Talk at the APS meeting(Washington,1977).
              M. Gell-Mann, P. Ramond and R. Slansky, Proceeding of supergravity 
              workshop at Stony Brook, eds. P. van Nieuwenhuisen and 
              D. Z. Freedman(North Holland, Amsterdam,1977).
%
\bibitem{cm}  S. Colemann and J. Mandula, {\it Phys. Rev.} {\bf 159}, 1251(1967).
%
\bibitem{hls} R. Haag, J. Lopuszanski, M. Sohnius, {\it Nucl. Phys.} {\bf B88}, 257(1975). 
%
\bibitem{ks1}  K. Shima, {\it Z. Phys.} {\bf C18}, 25 (1983).      
%
\bibitem{ks2}  K. Shima, {\it European. Phys. J.} {\bf C7}, 341(1999).
%
\bibitem{wb}  J. Wess and J. Bagger, {\it Supersymmetry and 
               Supergravity} (Princeton Univ.\ Press, 1992). 
%
\bibitem{ss}   A. Salam and J. Strathdee, \PL{B49} (1974) 465.
%
\bibitem{fwz}  S. Ferrara,  J. Wess and B. Zumino, \PL{B51}, 239(1974).
%
\bibitem{n}   W. Nahm, \NP{B135}, 149(1982).
%
\bibitem{fsz}  S. Ferrara,  C. A. Savoy and B. Zumino, \PL{B100}, 393(1981).
%
\bibitem{ks3}  K. Shima,  {\it Phys. Lett.} {\bf B501}, 237(2001).
%
\bibitem{bv}   W. Bardeen and V. Vi\v{s}nji\'{c}, \NP{B194}, 422(1982).
%
\bibitem{gg}  H. Georgi and S. Glashow, \PRL{32}, 438(1974). 
%
\bibitem{ks4}  K. Shima, {\it Fortschr. Physik}, {\bf 50}, 717(2002). \\
               Proceeding of the Symposium {\it 100 Years Werner Heisenberg-Works and Impact} 
               at Bamberg, Germany, 2001, eds. D. Papenfuss, D. Lust and W. Schleich(WILEY-VCH Verlag, Berlin, 2002).
%
\bibitem{dz1}  S. Deser and B. Zumino, \PRL{38} (1977) 1433.

%
\bibitem{ks0}  K. Shima, \PR{D20}, 574(1979). \\
               K. Shima, \PR{D15}, 216(1977). 
%
\bibitem{st1}  K. Shima and M. Tsuda, {\it Class. and Quantum Grav.} {\bf 19}, 1 (2002).
%
\bibitem{st2}  K. Shima and M. Tsuda, {\it Phys. Lett.} {\bf B507}, 260(2001).
%
\bibitem{ik}   E. A. Ivanov and A.A. Kapustnikov, {\it J. Phys.} {\bf A11}, 2375(1978).
%
\bibitem{r}    M. Ro\v{c}ek, {\it Phys. Rev. Lett.} {\bf 41}, 451(1978).
%
\bibitem{uz}   T. Uematsu and C.K. Zachos, \NP{B201} (1982) 250.
%
\bibitem{stt1}  K. Shima, Y. Tanii and M. Tsuda, {\it Phys. Lett.} {\bf B525}, 183(2002).

\bibitem{fi}    P. Fayet and J. Iliopoulos, \PL{B51} (1974) 461.
%
\bibitem{o}    L. O'Raifeartaigh, \NP{B96} (1975) 331.
%
\bibitem{f}    P. Fayet, \NP{B113} (1976) 135.
%
\bibitem{GSW}   N. Seiberg and E. Witten,  \NP{B426}, 19(1994). 
%
\bibitem{stt2}   K. Shima, Y. Tanii and M. Tsuda, {\it Phys. Lett.} {\bf B546}, 162 (2002).
%
\bibitem{GSW}   R. Grimm, M. Sohnius and J. Wess,  \NP{B133}, 275(1978). 

\bibitem{sts1}
K. Shima, M. Tsuda and M. Sawaguchi, {\it Czech. J. Phys.}{\bf38}, 21 (2002).
%
\bibitem{sts2}
K. Shima, M. Tsuda and M. Sawaguchi, hep-th/0211187(revised, 2003).   
Proceeding of the 4th Intrenational conference on symmetry in nonlinear mathematical physics, 
Kyiv, Ukraine(2003), eds. A. Nikitin, et al.
%
\bibitem{sts3} K. Shima, M. Tsuda and M. Sawaguchi, hep-th/0306080.   
%
%
\bibitem{sw}
K. Stelle and P. West,{\it Phys. Lett.} {\bf74}, 330(1978). 
%
\bibitem{fv}
S. Ferrara and P. van Nieuwenhuisen,{\it Phys. Lett.} {\bf74}, 333(1978). 
%
\bibitem{lr}
U. Lindstr\"om and M. Ro\v{c}ek, {\it Phys. Rev.}{\bf D19}, 2300 (1979).  
%
\bibitem{fr}
C. Fronsdal, {\it Phys. Rev.}{\bf D18}, 3624(1978).  
%
\bibitem{sts4}
K. Shima, M. Tsuda and M. Sawaguchi,  in preparation. 
%
\bibitem{b}  N.S. Baaklini, {\it Phys. Lett.} {\bf 67B}, 335(1976).
%
\bibitem{st3}
K. Shima and M. Tsuda, {\it Phys. Lett.} {\bf B521}, 67(2001).
%
\bibitem{s-ta}  K. Shima and  Y. Tanii, {\it Mod. Phys. Lett.}, {\bf A4}, 2259(1989). 
%
\bibitem{wttn} E. Witten,  {\it Commun. Math. Phys.}, {\bf80}, 381(1981).
%
\bibitem{bi}
M. Born and L. Infeld, {\it Proc. Roy. Soc.(London)}, {\bf144}, 425(1934). 

\end{thebibliography}
\end{document}